\documentclass[11pt,letterpaper]{article}

\usepackage[abbrvbib]{jmlr2e}

\usepackage[utf8]{inputenc}

\usepackage[table,dvipsnames]{xcolor}
\usepackage{epsfig}
\usepackage{pgfplotstable}
\usepackage{pgfplots}
\usepgfplotslibrary{groupplots}
\usepackage{amsmath}
\usepackage{amssymb}
\usepackage{bbm}
\usepackage{float}

\usepackage{microtype}
\usepackage{tikz}
\usetikzlibrary{decorations.text,calc,shapes,arrows,arrows.meta, positioning,shapes.misc,decorations.markings,decorations.markings,decorations.pathreplacing,matrix}
\usepackage{booktabs}
\usepackage{multirow}
\usepackage{adjustbox}
\usepackage{verbatim}
\usepackage[T1]{fontenc}
\usepackage[font={small,singlespacing},labelfont=it,labelsep=period]{caption}
\usepackage{adjustbox} %
\usepackage{graphicx}
\usepackage{color}
\usepackage{caption}
\usepackage{subcaption}
\usepackage[noend]{algpseudocode}
\usepackage{algorithm}

\makeatletter
\renewcommand{\ALG@beginalgorithmic}{\small}
\makeatother

\usepackage{siunitx} %
\usepackage[colorinlistoftodos,disable]{todonotes}

\DeclareMathOperator*{\argmin}{arg\,min}
\DeclareMathOperator*{\argmax}{arg\,max}

\pgfplotsset{
    discard if not/.style 2 args={
        x filter/.code={
            \edef\tempa{\thisrow{#1}}
            \edef\tempb{#2}
            \ifx\tempa\tempb
            \else
                
            \fi
        }
    }
}

\pgfplotsset{
    discard if nottwo/.style n args=4{
        x filter/.code={
            \edef\tempa{\thisrow{#1}}
            \edef\tempb{#2}
            \edef\tempc{\thisrow{#3}}
            \edef\tempd{#4}
            \ifx\tempa\tempb
              \ifx\tempc\tempd
              \else
                  
              \fi
            \else
                
            \fi
        }
    }
}

\makeatletter
\pgfplotsset{
    every axis x label/.append style={
        alias=current axis xlabel
    },
    legend pos/outer south/.style={
        /pgfplots/legend style={
            at={%
                (%
                \@ifundefined{pgf@sh@ns@current axis xlabel}%
                {xticklabel cs:0.5}%
                {current axis xlabel.south}%
                )%
            },
            anchor=north
        }
    }
}

\usepackage{xspace}

\newcommand\textcite[1]{\citet{#1}}
\renewcommand\cite[1]{\citep{#1}}

\jmlrheading{18}{2017}{1-5}{Submitted}{Published}{55-555}{Edward Raff}

\ShortHeadings{
Toward Metric Indexes for Incremental Insertion and Querying
}{
Raff and Nicholas
}
\firstpageno{1}

\begin{document}

\def\rbci{$\text{RBC}_{\text{Imp}}$\xspace}
\def\vpmv{$\text{VP}_\text{MV}$\xspace}
\def\coverb{$\text{Cover}_\text{B}$\xspace}
\def\coveri{$\text{Cover}_\text{I}$\xspace}

\title{Toward Metric Indexes for Incremental Insertion and Querying}

\author{\name Edward Raff \email edraff@lps.umd.edu \\
\addr Laboratory for Physical Sciences\\ 
Booz Allen Hamilton \\
University of Maryland, Baltimore County
\AND 
\name Charles Nicholas \email nicholas@umbc.edu \\
\addr University of Maryland, Baltimore County
	}

\editor{}

\maketitle

\begin{abstract}
In this work we explore the use of metric index structures, which accelerate nearest neighbor queries, in the scenario where we need to interleave insertions and queries during deployment. This use-case is inspired by a real-life need in malware analysis triage, and is surprisingly understudied. Existing literature tends to either focus on only final query efficiency, often does not support incremental insertion, or does not support arbitrary distance metrics.  We modify and improve three algorithms to support our scenario of incremental insertion and querying with arbitrary metrics, and evaluate them on multiple datasets and distance metrics while varying the value of $k$ for the desired number of nearest neighbors. In doing so we determine that our improved Vantage-Point tree of Minimum-Variance performs best for this scenario. 
\end{abstract}

\begin{keywords}
  nearest neighbor, incremental, search, metric index, metric space.
\end{keywords}

\section{Introduction}

Many applications are built on top of distance metrics and nearest neighbor queries, and have %
achieved better performance through the use of metric indexes. A metric index is a data structure used to answer neighbor queries that accelerates these queries by avoiding unnecessary distance computations. The indexes we will look at in this work require the use of a valid distance metric (i.e., obeys triangle inequality, symmetry, and indiscernibility) and returns exact results. 

Such indexes can be used to accelerate basic classification and similarity search, as well as many popular clustering algorithms like k-Means \cite{Lloyd1982,Kanungo2002}, density based clustering algorithms like DBSCAN \cite{Bicici2007,hdbscan}, and visualization algorithms like t-SNE \cite{VanderMaaten2014,Maaten2008,Tang:2016:VLH:2872427.2883041,Narayan2015}. However, most works assume that the data to be indexed is static, and that there will be no need to update the index over time. Even when algorithms are developed with incremental updates, the evaluation of such methods is not done in such a context. In this work we seek to evaluate metric indexes for the case of incremental insertion and querying. Because these methods are not readily available, we modify three existing indexes to support incremental insertion and querying. 

Our interest in this area is particularly motivated by an application in malware analysis, where we maintain a database of known malware of interest. Malware may be inserted into the database with information about malware type, method of execution, suspected origin, or suspected author. When an analyst is given new malware to dissect, the process can be made more efficient if a similar malware sample has already been processed, and so we want to efficiently query the database to retrieve potentially related binaries. This triaging task is a common problem in malware analysis, often related to malware family detection \cite{178990,Gove:2014:SSV:2671491.2671496,walenstein2007exploiting,Jang2011}. Once done, the analyst may decide the binary should be added to the database. In this situation our index would be built once, and have insertions into the database regularly intermixed with queries. This read/write ratio may depend on workload, but is unfortunately not supported by current index structures that support arbitrary distance metrics. This scenario inspires our work to build and develop such indexes, which we test on a wider array of problems than just malware. We do this in part because the feature representations that are informative for malware analysis may change, along with the distance metrics used, and so a system that works with a wide variety of distance measures is appropriate. 

To emphasize the importance of such malware triage, we note it is critical from a time saving perspective. Such analysis requires extensive expertise, and it take an expert analysis upward of 10 hours to dissect a single binary \cite{Mohaisen:2013:UZA:2487788.2488056}. Being able to identify a related binary that has been previously analyzed may yield significant time savings.  The scale of this problem is also significant. A recent study of 100 million computers found that 94\% of files were unique \cite{Li2017}, meaning exact hashing approaches such as MD5 sums will not help, and similarity measures between files are necessary. In terms of incremental addition of files, in 2014 most \todo[color=green]{\footnotesize AV? -CNK, Yes -Ed} anti-virus vendors were adding 2 to 3 million new binaries each month \cite{tagkey2014iv}. 

Given our motivation, we will review the related work to our own in \autoref{sec:related_work}. We will review and modify three algorithms for incremental insertion and querying in \autoref{sec:metrics_used}, followed by the evaluation details, datasets and distance metrics in \autoref{sec:data_method}. Evaluations of our modifications and their impact will be done in \autoref{sec:eval_nn_prune_improve}, followed by an evaluation of the incremental insertion and querying scenario in \autoref{sec:eval_inc_const}. Finally, we will present our conclusions in \autoref{sec:conclusions}.

\section{Related Work} \label{sec:related_work}

There has been considerable work in general for retrieval methods based on $k$ nearest neighbor queries, and many of the earlier works in this area did support incremental insertion and querying, but did not support arbitrary distance metrics. One of the earliest methods was the Quad-Tree\cite{Finkel:1974:QTD:2697709.2697865}, which was limited to two-dimensional data. This was quickly extended with the kd-tree, which also supported insertions, but additionally supported arbitrary dimensions and deletions as well\cite{Bentley:1975:MBS:361002.361007}. However, the kd-tree did not support arbitrary metrics, and was limited to the euclidean and similar distances. Similar work was done for the creation of R-trees, which supported the insertion and querying of shapes, and updating the index should an entry's shape change\cite{Guttman1984}. However improving the query performance of R-trees involved inserting points in a specific order, which requires having the whole dataset available from the onset\cite{Kamel:1994:HRI:645920.673001}, and still did not support arbitrary metrics. 

The popular ball-tree algorithm was one of the first efforts to devise and evaluate multiple construction schemes, some which required all the data to be available at the onset, while others which could be done incrementally as data became available \cite{Omohundro1989}. This is similar to our work in that we devise new incremental insertion strategies for two algorithms, though \citet{Omohundro1989} do not evaluate incremental insertions and querying. This ball-tree approach was limited to the euclidean distance primarily from the use of a mean data-point computed at every node. Other early work that used the triangle inequality to avoid distance computations had this same limitation \cite{Fukunage1975}. 

\todo[inline,color=yellow]{what IS a metric index? -CNK. Added definition at intro. Good? -Ed}
While almost all of these early works in metric indexes  supported incremental insertion, none contain evaluation of the indexes under the assumption of interleaved insertions and queries. These works also do not support arbitrary distance metrics. 

The first algorithm for arbitrary metrics was the metric-tree structure \cite{Uhlmann1991,Uhlmann1991a}, which used the distance to a randomly selected point to create a binary tree. This was independently developed, slightly extended, and more throughly evaluated to become the Vantage-Point tree we explore in this work\cite{Yianilos1993}. However, these methods did not support incremental insertion. We will modify and further improve the Vantage-Point tree in \autoref{sec:metrics_used}. 

Toward the creation of provable bounds for arbitrary distance metrics, the concept of the expansion constant $c$ was made by \citet{Karger:2002:FNN:509907.510013}. The expansion constant is a property of the current dataset under a given metric, and describes a linear relationship between the radius around a point, and the number of points contained within that radius. That is to say, if the radius from any arbitrary point doubles, the number of points contained within that radius should increase by at most a constant factor. Two of the algorithms we look at in this work, as discussed in \autoref{sec:metrics_used}, make use of this property. 

The first practical algorithm to make use of the expansion constant was the Cover-tree \cite{Beygelzimer2006}, which showed practical speed-ups across multiple datasets and values of $k \in [1, 10]$. Their results were generally shown under $L_p$ norm distances, but also included an experiment using the string edit distance. Later work then simplified the Cover-tree algorithm and improved performance, demonstrating its benefit on a wider variety of dataset and distance metrics \cite{Izbicki2015}. Of the algorithms for metric indexes, the Cover-tree is the only one we are aware of with an incremental construction approach, and so we consider it one of our metrics of interest in \autoref{sec:metrics_used}. While the Cover-tree construction algorithm is described as an incremental insertion process, the more efficient variant proposed by \citet{Izbicki2015} includes a bound which requires the whole dataset in advance to calculate bounds, preventing the efficient interleaving of insertions and queries\footnote{The original Cover-tree did not have this issue, and so would meet our requirements for incremental insertion. We consider the newer variant since it is the most efficient.}.  

Another algorithm we consider is the Random Ball Cover (RBC), which was designed for making effective use of GPUs with the euclidean distance \cite{Cayton2012}. Despite testing on only the euclidean distance, the algorithm and proof does not rely on this assumption -- and will work with any arbitrary distance metric. We consider the RBC in this work due to its random construction, which allows us to devise an incremental construction procedure that closely matches the original design and maintains the same performance characteristics. While the Random Ball Cover has inspired a number of GPU based follow ups \cite{Li:2015:BKN:2977215.2977244,Kim2013,Gieseke:2014:BKT:3044805.3044826}, we do not assume that a GPU will be used in our work. 

\citet{Li2016} develop an indexing scheme that supports incremental updates, but only works for the euclidean distance. They also do not evaluate the performance as insertions and queries are interleaved. 

\section{Metric Indexes Used} \label{sec:metrics_used}

\todo[inline,color=yellow]{should it be indices or indexes? -CNK I think both are considered valid - Ed}
Given the existing literature of metric indexes there appear to be no readily available methods that suit our needs. For this reason we take three algorithms and modify them for incremental index construction and querying. In particular, we adapt the Random Ball Cover, Vantage Point tree, and Cover-tree algorithms for incremental insertion. As classically presented, the first two methods methods are not designed for this use case. While the original cover tree algorithm did support incremental insertions, its improved variants do not. More importantly, as we will show in \autoref{sec:eval_nn_prune_improve}, the Cover-tree has worse than brute-force performance with one of our distance metrics.  With our modifications we satisfy three goals that have not yet been achieved in a single data structure:

\begin{enumerate}
\item New datapoints can be added to the index at any point
\item We can efficiently query the index after every insertion
\item The index can be efficiently used with any distance metric
\end{enumerate}

\begin{figure}[!htb]
\centering
\begin{subfigure}[t]{.32\textwidth}
  \centering
  \resizebox{\textwidth}{!}{%
  	\includegraphics{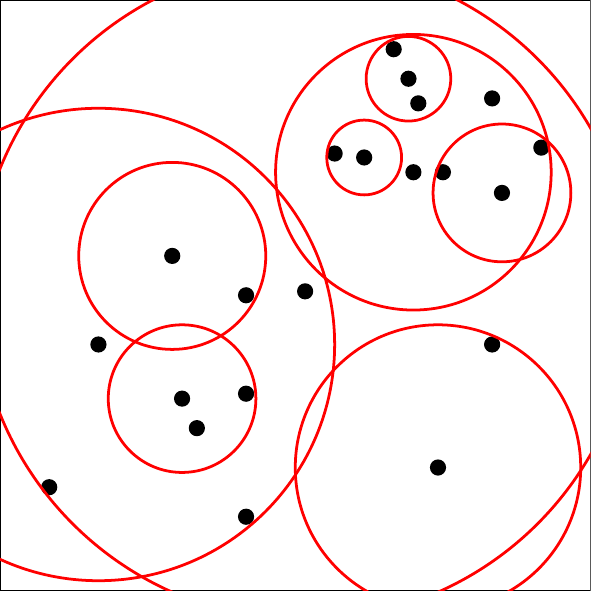}
  }
  \caption{\footnotesize Cover-trees produce a heiarchy of circles, but each node may have a variable number of children. Each node has a radius that upper bounds the distance to all of its children, and may partially overlap.}
  \label{fig:cover_partition_example}
\end{subfigure}\hfill %
\begin{subfigure}[t]{.32\textwidth}
  \centering
  \resizebox{\textwidth}{!}{%
  	\includegraphics{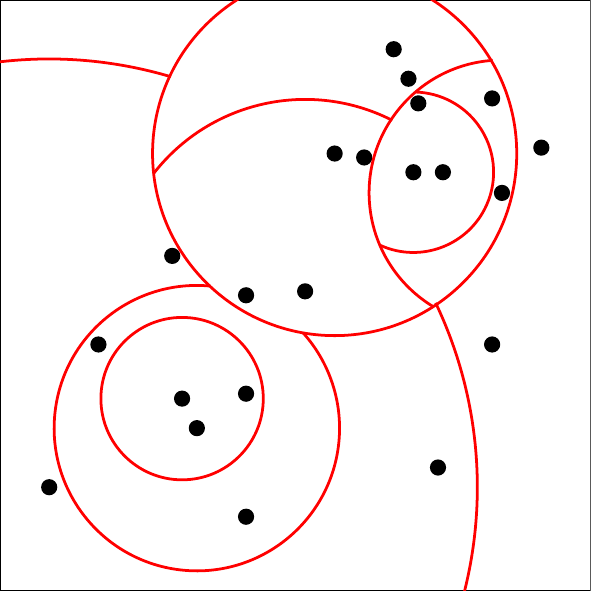}
  }
  \caption{\footnotesize Vantage-Point trees divide the space using a hierarchy of circles. The in/outside of each space acts as a hard boundary when subdividing.}
  \label{fig:vp_partition_exampled}
\end{subfigure}\hfill%
\begin{subfigure}[t]{.32\textwidth}
  \centering
  \resizebox{\textwidth}{!}{%
  	\includegraphics{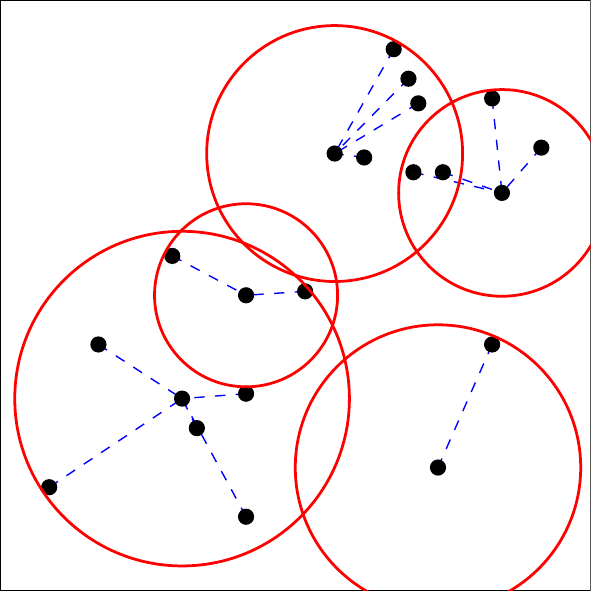}
  }
  \caption{\footnotesize  RBC selects a subset of representatives, and each point is assigned to its nearest representative (relationships marked with dashed blue line).}
  \label{fig:rbc_partition_example}
\end{subfigure}
\caption{Example partitionings for all three algorithms. Red circles indicate the radius from which one node covers out in the space.  }
\label{fig:test}
\end{figure}

While the latter point would seem satisfied by the original Cover-tree algorithm, our results indicate a degenerate case where the Cover-tree performs significantly worse than a brute force search. For this reason we consider it to have not satisfied our goals. 

We also contribute improvements to both the Random Ball Cover and Vantage Point Tree structures that further reduce the number distance computations needed by improving the rate at which points are pruned out. These improvements can dramatically increase their effective pruning rate, which leads us to alter our conclusions about which method should be used in the general case. 

In the below descriptions, we will use $S$ to refer to the set of points currently in the index, and $n = |S|$ as the number of such points. A full review of all details related to the three methods is beyond this scope of this work, but we will provide the details necessary to understand what our contributions are to each approach. 

\subsection{Cover Tree} \label{sec:cover_tree}

The Cover-tree \cite{Beygelzimer2006} is a popular method for accelerating nearest neighbor queries, and one of the first practical metric indexes to have a provable bound using the expansion constant $c$ \cite{Karger:2002:FNN:509907.510013}. The Cover-tree can be constructed in $O(c^6 n \log n)$ time, and answer queries in $O(c^{12} \log n)$ time. \textcite{Izbicki2015} developed the Simplified Cover Tree, which reduces the practical implementation details and increases efficiency in both runtime and avoiding distance computations.\footnote{\citeauthor{Izbicki2015} also introduced a Nearest Ancestor Cover Tree, but we were unable to replicate these results. The reported performance difference between these two variants was not generally large, and so we use only the simplified variant.} To reproduce the Simplified Cover Tree algorithm without any nearest-neighbor errors, we had to make two slight modifications
to the algorithm as originally presented. These adjustments are detailed in  \autoref{sec:cover_correction}.

The Cover-tree algorithm, as its name suggests, stores the data as a tree structure where each node represents only one data point and may have any number of children nodes\footnote{The maximum number of children is actually bounded by the expansion constant $c$.}. The tree is constructed via incremental insertions, which means we require no modifications to the construction algorithm to support our use case. However, at query time it is necessary for each node $p$ in the tree to compute a \textit{maxdist}, which is the maximum distance from the point represented by node $p$ to any of its descendant nodes. This maxdist value is used at every level of the tree to prune children nodes from the search path. Insertions can cause re-organizations of the tree, resulting in the need to re-compute maxdist bounds. For this reason the Simplified Cover-tree can not be used to efficiently query the index between consecutive insertions. 

Because of the re-balancing and re-organization that occurs during tree construction, it is not trivial to selectively update the maxdist value based on the changes that have occurred. Instead we will use an upper bound on the value of maxdist. Each node in the tree maintains a maximum child radius of the form $2^l$, where $l$ is an integer. This also upper bounds the maxdist value of any node by $2^{l+1}$ \cite{Izbicki2015}. This will allow us to answer queries without having to update maxdist, but results in a loosening of the bound. The performance of this upper bounded version of the Cover-tree we will refer to as \coverb, and is more naturally suited to the use case of interleaved insertions and queries. 

We note as well that this relaxation on the maxdist based bound represents a compromise between the simplified approach proposed by \citeauthor{Izbicki2015} and the original formulation by \citeauthor{Beygelzimer2006}. In the later case, the $2^{l+1}$ bound is used to prune branches, but all branches are traversed simultaneously. In the former, the maxdist bound is used to descend the tree one branch at a time, and the nearest neighbor found so far is used to prune out new branches. By replacing maxdist with $2^{l+1}$, we fall somewhere in-between the approaches. Using a looser bound to prune, but still avoiding traversing all branches. In our extensive tests of these algorithms, we discovered two issues with the original specification of the simplified Cover-tree. These are detailed in \autoref{sec:cover_correction}, along with our modifications that restore the Cover-tree's intended behavior.

\subsection{Vantage Point Tree} \label{sec:vp_tree}

\begin{figure}[!htb]
\centering
\begin{tikzpicture}
\coordinate (O) at (0,0);

\draw[orange,dotted,line width=0.5mm] (O) circle (2.5);
\draw[OliveGreen,dotted,line width=0.5mm] (O) circle (1.75);
\draw[blue,dotted,line width=0.5mm] (O) circle (1.5);
\draw[red,dotted,line width=0.5mm] (O) circle (0.5);
\draw[fill] (O) circle (0.075);

\draw[fill] (0, 0.5) circle (0.075);
\draw[fill] (0.6, 0.54) circle (0.075);
\draw[fill] (-0.75, 0.9) circle (0.075);
\draw[fill] (-0.65, -0.55) circle (0.075);
\draw[fill] (-1.5, 0) circle (0.075);
\draw[fill] (0, -1.75) circle (0.075);
\draw[fill] (0.9, 1.94) circle (0.075);
\draw[fill] (1.95, -1.25) circle (0.075);
\draw[fill] (-1.75, -1.45) circle (0.075);
\draw[fill] (2.5, 0) circle (0.075);

\end{tikzpicture}
\caption{Example of a node in a vp-tree, with the vantage point in the center. The \textcolor{red}{low-near} bound is in \textcolor{red}{red}, the distance to the point closest to the center. The \textcolor{blue}{low-far} (\textcolor{blue}{blue}) and  \textcolor{OliveGreen}{high-near} (\textcolor{OliveGreen}{green}) braket the boundry of the median. No points can fall between these bounds. The farthest away point provides the \textcolor{orange}{high-far} bound in \textcolor{orange}{orange}. }
\label{fig:vp_low_high_bounds}
\end{figure}
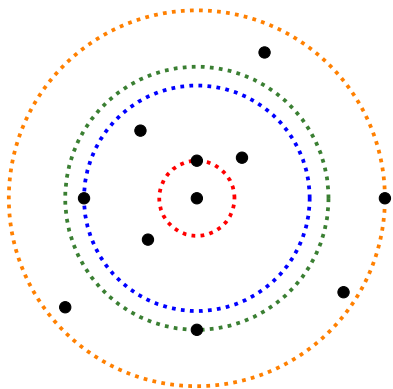

The Vantage Point tree \cite{Yianilos1993,Uhlmann1991} (VP-tree) is one of the first data structures proposed for accelerating neighbor searches using an arbitrary distance metric. The construction of the VP-tree results in a binary tree, where each node $p$ represents one point from the dataset, the "vantage point". The vantage point splits its descendant into a low and high range based on their distance from the aforementioned vantage point, with half of the child vectors in each range. For each range, we also have a nearest and farthest value, and an example of how these are used is given in \autoref{fig:vp_low_high_bounds}. 

This tree structure is built top-down, and iteratively splits the remaining points into two groups at each node in the tree. Rather than continue splitting until each node has no children, there is instead a minimum split size $b$. This is because there are likely too few points for which we can obtain good low/high bounds. Instead, once the number of datapoints is $\leq b$, we create a "bucket" leaf node that stores the points together and uses the distance from each point to its parent node to do additional pruning. 

At construction time, since each split is done by breaking the tree in half, the maximum depth of the tree is $O(\log n)$ and construction takes $O(n \log n)$ time. Assuming the bounds are successful in pruning most branches, the VP-tree then answers queries in $O(\log n)$ time. 

The bucketing behavior can provide practical runtime performance improvements as well. Some of this comes from better caching behavior, as bucket values will be accessed in a sequential pattern, and avoids search branches that can be more difficult to accurately predict for hardware with speculative execution. This can be done for the VP-tree because its structure is static as it is created, where the Cover-tree cannot create bucket nodes due to the re-balancing done during construction.

\subsubsection{Incremental Construction} \label{sec:vp_inc_construction}

While the Cover-tree required minimal changes since its construction is already incremental, we must define a new method to support such a style for the VP-tree. To support incremental insertions into a VP-tree, we must first find a location with which to store the new datapoint $x$. This can be done quite easily by descending the tree via the low/high bounds stored for each point, and updating the bounds as we make the traversal. One we reach a leaf node, $x$ is simply inserted into the bucket list. However, we do not expand the leaf node when its size exceeds $b$. 

Ideally, these bounds will be changed infrequently as we insert new points. Getting a better estimate of the initial bound values should minimize this occurrence. For this reason we expand a bucket $b$ once it reaches a size of $b^2$. This gives us a larger sample size with which to estimate the four bound values. We use the value $b^2$ as a simple heuristic that follows our intuition that a larger sample is needed for better estimates, allows us to maintain the fast construction time of the VP algorithm, and results in an easy to implement and replicate procedure. 

\begin{algorithm}[!ht]
\caption{Insert into VP-tree}
\label{algo:vp_insert}
\begin{algorithmic}[1]
\Require vp-tree root node $p$, and new datapoint $x$ to insert into tree.
\While{$p$ is not a leaf node}
	\State $dist \gets d(x, p.vp)$
	\If{$dist < (p.\text{low}_{\text{far}} + p.\text{high}_{\text{near}})/2$}
    	\State $p.\text{low}_{\text{far}} \gets \max\left(dist, p.\text{low}_{\text{far}}\right)$
        \State $p.\text{low}_{\text{near}} \gets \min\left(dist, p.\text{low}_{\text{near}}\right)$
    	\State $p \gets p.\text{lowChild}$
    \Else
    	\State $p.\text{high}_{\text{far}} \gets \max\left(dist, p.\text{high}_{\text{far}}\right)$
        \State $p.\text{high}_{\text{near}} \gets \min\left(dist, p.\text{high}_{\text{near}}\right)$
    	\State $p \gets p.\text{highChild}$
    \EndIf
\EndWhile
\State Add $x$ to bucket leaf node $p$
\If{$|p.\text{bucket}| > b^2$ }
	\State Select vantage point from $p.\text{bucket}$ and create a new split, adding two children nodes to $p$. 
\EndIf
\State \Return
\end{algorithmic}
\end{algorithm}  

Thus our insertion procedure is given in \autoref{algo:vp_insert}, and is relatively simple. Assuming the tree remains relatively balanced, we will have an insertion time of $O(\log n)$. This will also maintain the query time of $O(\log n)$.

\subsubsection{Faster Search}

We also introduce a new modification to the VP-tree construction procedure that reduces search time by enhancing the ability of the standard VP-tree search procedure to prune out branches of the tree. This is done by using an extension of the insight from \autoref{sec:vp_inc_construction}, that we want to make our splits only when we have enough information to do so. That is, once we have enough data to make a split, choosing the median distance from the vantage point may not be the smartest split. 

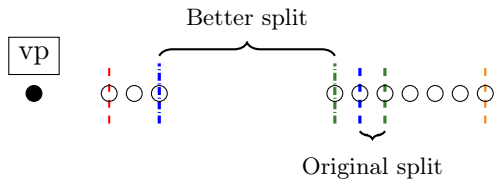
\begin{figure}[!htb]
\centering
\begin{tikzpicture}
	\draw[red, dashed, thick]    (1,-0.4) -- (1,0.4);
    \draw[orange, dashed, thick] (6,-0.4) -- (6,0.4);
    \draw[thick,decoration={brace,mirror,amplitude=3pt},decorate]
    	(4.333,-0.5) -- node[below=6pt] {\footnotesize Original split} (4.666,-0.5);
    \draw[blue, dashed, very thick]    (4.333,-0.4) -- (4.333,0.4);
    \draw[OliveGreen, dashed, very thick]   (4.666,-0.4) -- (4.666,0.4);
    \draw[thick,decoration={brace,amplitude=5pt},decorate]
    	(1.666,0.5) -- node[above=6pt] {\footnotesize Better split} (4,0.5);
    \draw[blue, densely dashdotted, very thick]    (1.666,-0.4) -- (1.666,0.4);
    \draw[OliveGreen, densely dashdotted, very thick]   (4,-0.4) -- (4,0.4);

	\draw[fill] (0,0) circle (3pt);
	\foreach \i in {0,...,2}
	{
		\draw (1+\i/3,0) circle (3pt);
	}
    \foreach \i in {0,...,6}
	{
		\draw (4+\i/3,0) circle (3pt);
	}

    \node[draw,align=center] at (0,0.5) {vp};
\end{tikzpicture}
\caption{Example on how the split can be improved, with vantage point in black and other points sorted by distance to it. Colors correspond to \autoref{fig:vp_low_high_bounds}. }
\label{fig:vp_better_split}
\end{figure}

Instead, we can use the distribution of points from the vantage point to choose a split that better bifurcates the data based on the distribution. An example of this is given in \autoref{fig:vp_better_split}, where the data may naturally form a binary split. This increases the gap between the $\text{low}_{\text{far}}$ and $\text{high}_{\text{near}}$ bounds, which then allows the search procedure to more easily prune one of the branches. 

To do this quickly, so to minimize any increase in construction time, we borrow from the CART algorithm used to construct a regression tree\cite{Breiman1984}. Given a set of $n$ distances to the vantage-point, we find the split that minimizes the weighted variance of each split

\begin{equation}\label{eq:vairance_split}
\argmin_s s \cdot \sigma_{1:s}^2 + (n-s) \cdot \sigma_{s:n}^2
\end{equation}

Where  $\sigma_{s:n}^2$ indicates the variance of the points in the range of $[s, n)$ when sorted by distance to the vantage point.  Because \eqref{eq:vairance_split} can be solved with just two passes over the $n$ points~\cite{Welford1962a,Chan1983}, we can solve this quickly with only an incremental increase in runtime. 

The original VP tree selects the median distance of all points from the vantage point. This requires $n$ distance computations, and an $O(n)$ quick-select search. Finding the split of median variance still requires $n$ distance computations, so that cost remains unchanged. However, a sort of $O(n \log n)$ must be done to find the split of minimum variance. 

\subsection{Random Ball Cover} \label{sec:rbc}

The Random Ball Cover~\cite{Cayton2012} (RBC) algorithm was originally proposed as an accelerating index that would make efficient use of many-core systems, such as GPUs. 
This was motivated by the euclidean distance metric, which can be computed with high efficiency when computing multiple distances simultaneously. This can be done by exploiting a decomposition of the euclidean distance into matrix operations, for which optimized BLAS routines are readily available. To exploit batch processing while also pruning distances, the RBC approach organizes data into large groups and uses the triangle inequality sparingly to prune out whole groups at a time. Compared to the VP and Cover Tree, the RBC algorithm is unique in that it aims to answer queries in $O(\sqrt{n})$ time and perform construction in $O(n \sqrt{n})$ time. 

The training procedure of the RBC algorithm is to randomly select $O(\sqrt{n})$ centers from the dataset, and denote that set of points as $R$. These are the $R$ \textit{random balls} of the algorithm. Each representative  $r_i \in R$  will own, or \textit{cover}, all the datapoints for which it is the nearest neighbor, $\argmin_{x} d(x, r_i) \forall x \in S \setminus R$, which is denoted as $L_{r_i}$. It is expected that each $r_i$ will then own $O(\sqrt{n})$ datapoints. Querying is done first against the subset of points $R$, from which many of the representatives are pruned. Then a second query is done against the points owned by the non-pruned representatives. To do this pruning, we need the representatives to be sorted by their distance to the query point $q$. We will denote this as $r_i^{(q)}$, which would be the $i$'th nearest representative to $q$. Pruning for $k$ nearest neighbor queries is then done using two bounds, 

\begin{equation}\label{eq:rbc_b1}
d(q, r_i) < d(q, r_k^{(q)}) + \psi_{r_i}
\end{equation}

\begin{equation}\label{eq:rbc_b2}
d(q, r_i) < 3 \cdot d(q, r_k^{(q)})
\end{equation}

Where $\psi_{r_i} = \max_{x \in L_{r_i}} d(r_i, x)$ is the radius of each representative, such that all datapoints fall within that radius. Each bound must be true for any $r_i$ to have the $k$'th nearest neighbor to query $q$, and the overall procedure is given in  \autoref{algo:rbc_search_orig}. Theoretically the RBC bounds are interesting in that they provide a small dependency on the expansion constant $c$ of the data, where queries can be answered in $O(c^{3/2} \sqrt{n})$ time. This is considerably smaller than the $c^{12}$ term in cover trees, but has the larger $\sqrt{n}$ dependence \todo[color=yellow]{right word? -CNK I think so -Ed} on $n$ instead of logarithmic. However, the RBC proof depends on setting the number of representatives $|R| = O(c^{3/2} \sqrt{n})$ as well, which we would not know in advance in practice. Instead we will use $|R| = \sqrt{n}$ in all experiments. 

\begin{algorithm}[!htb]
\caption{Original RBC Search Procedure}
\label{algo:rbc_search_orig}
\begin{algorithmic}[1]
\Require Query $q$, desired number of neighbors $k$
\State Compute sorted order $r_i^{(q)}$ $ \forall r \in R$ by $d(r, q)$ 
\State FinalList $\gets \emptyset$

\ForAll{$r_i \in R$ } 
	\If {Bounds \eqref{eq:rbc_b1} and \eqref{eq:rbc_b2} are True}
    	\State FinalList $\gets$ FinalList $\cup$  $L_{r_i}$
    \EndIf
\EndFor
\State $k$-NN $\gets$ \Call{BruteForceSearch}{$q$, $R$ $\cup$ FinalList} \Comment{distances for $R$ do not need to be re-computed}
\State \Return $k$-NN
\end{algorithmic}
\end{algorithm}  

\subsubsection{Incremental Construction} \label{sec:rbc_inc_construction}

If our goal was to build a static index, the random selection of $R$ may lead to a sub-optimal selection. It is possible that different representatives will have widely varying numbers of members. For our goal of incrementally adding to an index, this stochastic construction becomes a benefit. Because the representatives are selected randomly without replacement, it is possible to incrementally add to the RBC index while maintaining the same quality of results. 

\begin{algorithm}[!htb]
\caption{Insert into RBC Index}
\label{algo:rbc_insert}
\begin{algorithmic}[1]
\Require RBC representatives $R$, associated lists $L_r, \forall r \in R$, and new datapoint $x$ to add to RBC.
\State Compute sorted order $r_i^{(x)}$ $ \forall r \in R$ by $d(r, x)$ 
\State $L_{r_1^{(x)}} \gets L_{r_1^{(x)}} \cup x$
\State $\psi_{r_1^{(x)}} \gets \max\left(d(r_1^{(x)}, x) ,\psi_{r_1^{(x)}} \right) $ \Comment{keep radius information correct}
\If {$\text{ceil}\left(\sqrt{n}\right)^2 \neq n$} 
	\State \Return \Comment{else, expand R set}
\EndIf
\State select randomly a datapoint $l_{\text{new}}$ from $\bigcup_{\forall r \in R} L_r$ 
\State let $r_{old}$ be the representative that owns $l_{\text{new}}$, i.e., $l_{\text{new}} \in L_{r_{old}}$
\State $L_{r_{old}} \gets L_{r_{old}} \setminus l_{\text{new}} $
\State $r_{\text{new}} \gets l_{\text{new}} $
\State potentialChildren $\gets$ \Call{RadiusSearchRBC}{$r_{\text{new}}$, $\argmax_{r, \forall r \in R} \psi_r$}
\State $L_{r_{\text{new}}} \gets \emptyset $
\State $R \gets R \cup r_{\text{new}}$
\State $\psi_{r_{\text{new}}} \gets 0$
\ForAll { $y \in $ potentialChildren }
	\State Let $r_y$ be the representative that owns $y$
    \If {$d(y, r_y) > d(y, r_{\text{new}})$} \Comment{change ownership}
    	\State $L_{r_{y}} \gets L_{r_{y}} \setminus y $
        \State $L_{r_{new}} \gets L_{r_{new}} \cup y $
        \State $\psi_{r_{y}} \gets\argmax_{\forall z \in L_{r_y}} d(r_y, z)  $ \Comment{update radius info}
        \State $\psi_{r_{\text{new}}} \gets \max \left(\psi_{r_{\text{new}}}, d(y, r_{\text{new}}) \right)$
    \EndIf
\EndFor

\end{algorithmic}
\end{algorithm}  

The details of our approach are given in \autoref{algo:rbc_insert}. Whenever we add a new datapoint to the index, we find its representative and add it to the appropriate list $L$. This can be done in $O(\sqrt{n})$ time, consistent with the query time of RBC. Once the closest representative is found, the radius to the farthest point may need to be updated, which is trivial. For the majority ($n-\sqrt{n}$) of insertions, this is all the work that needs to be done. 

For the remaining $\sqrt{n}$ insertions, the total number of datapoints will reach a size such that we should have a new representative. The new representative will be selected randomly from all the points in $S \setminus R$. We can find the all the datapoints that may belong to this new representative using a "range" or "radius" search. A radius search is given a query and radius, and returns all datapoints within the specified radius of the query. In this case we give the new representative as the query and specify the range as the maximum $\psi_r$ in the RBC so far. This is by definition the maximum distance of any point to its representative, so any point that will be owned by the new representative must have a smaller distance. 
In the worst case scenario, we cannot prune any points using a radius search. This means at most $n$ other points must be considered. But since this scenario can only occur $\sqrt{n}$ times, we maintain the same construction time complexity of $O(n \sqrt{n})$ in all cases. We can also state that this approach yields an amortized $O^*(\sqrt{n})$ insertion time. 

\subsubsection{Faster Search}

While the original RBC search is fast and efficient on GPUs and similar many-core machines, it is not as efficient for our use case. Our scenario of interleaved insertions and queries means will be querying with only a few datapoints at a time. This means we will not obtain a large enough group of queries points to obtain the batch and SIMD efficiencies that were the original goal of \citet{Cayton2012}. Further, when we consider arbitrary distance metrics, we can not expect the same efficient method of grouping calculations as can be done with the euclidean distance. Thus we have developed an improved querying method for the RBC search to make it  more efficient in our incremental insertion and querying scenario. Our improvements to the RBC search procedure can be broken down into three steps.  

First, we modify the search to create the $k$-NN list incrementally as we visit each representative $r \in R$. In particular we can improve the application of bound \eqref{eq:rbc_b1} by doing this. First, we note that in \eqref{eq:rbc_b1}, the $d(q, r_k^{(q)})$ term serves as an upper bound on the distance to the $k$'th nearest neighbor. By building the $k$-NN list incrementally, we can instead use the current best candidate for $k$'th nearest neighbor as a bound on the distance to the $k$'th nearest neighbor. This works intuitively, as the true $k$'th neighbor, if not yet found, must by definition have a smaller distance than our current candidate. 

Second, when visiting the points owned by each representative, $l \in L_r$, we can apply this bound again and tighten the bound further. This is done by replacing the $\psi_{r_i}$ term of \eqref{eq:rbc_b1} by the distance of $l$ to its representative $r$. Since this distance $d(l, r)$ had to be computed when building the RBC in the first place, these distances can simply be cached at construction --- avoiding any additional overhead. 

Third, to increase the likelihood of finding the $k$'th neighbor earlier in the process, we visit the representatives in sorted order by their distance to the query. Because our first modification tightens the bound as we find better $k$'th candidates, this will accelerate the rate at which we tighten the bound. 

The complete updated procedure is given in \autoref{algo:rbc_search_new}. A similar treatment can improve the RBC search procedure for range queries. We note that one lines 2 through 4, we add all the children points of the closest representative $L_{r_1^{(q)}}$ unconditionally. This satisfies requirements of the RBC search algorithm's correctness in the $k$ nearest neighbor case, rather than just one nearest neighbor. We refer the reader to \citet{Cayton2012} for details. The essence of its purposes is to pre-populate the $k$-NN list with values for the bounds checks done in lines 8 and 10.

\begin{algorithm}[!htb]
\caption{New RBC Search Procedure}
\label{algo:rbc_search_new}
\begin{algorithmic}[1]
\Require Query $q$, desired number of neighbors $k$
\State Compute sorted order $r_i^{(q)}$ $ \forall r \in R$ by $d(r, q)$ 
\State $k$-NN $\gets \{r_1^{(q)}\}$ \Comment{sorted list implicitly maintains max size of $k$}
\ForAll{$l \in L_{r_1^{(q)}}$} \Comment{Add the children of the nearest representative}
	\State $k$-NN $\gets$ $k$-NN $\cup$ $l$
\EndFor
\For{$i \in 2 \ldots |R|$  } \Comment{visit representatives in sorted order}
	\State $qr \gets d(q, r_i^{(q)})$
    \State Add tuple $r_i^{(q)}$, $d(r_i^{(q)}, q)$ to $k$-NN
	\If {$qr < k$-NN[k].dist  $+$ $\psi_{r_i}$ and \eqref{eq:rbc_b2} are True}
    	\ForAll {$l \in L_{r_i^{(q)}}$ } 
        	\If {$qr < k$-NN[k].dist  $+$ $d(l, r_i^{(q)})$} \Comment{$d(l, r_i^{(q)})$ is pre-computed}
            	\State Add tuple $l$, $d(l, q)$ to $k$-NN
            \EndIf
        \EndFor
    \EndIf
\EndFor
\State \Return $k$-NN
\end{algorithmic}
\end{algorithm}

The first step of our new algorithm must still compute the distances for each $r_i$, and $|R| = \sqrt{n}$. In addition, we add all the children of the closest represent $r_1^{(q)}$, which is expected to own $O(\sqrt{n})$ points.  Thus this modified RBC search is still an $O(\sqrt{n})$ search algorithm. Our work does not improve the algorithmic complexity but does improve its effectiveness at pruning. 

\section{Datasets and Methodology} \label{sec:data_method}

We use a number of datasets and distance metrics to evaluate our changes and the efficiency of our incremental addition strategies. For all methods we have confirmed that the correct nearest neighbors are returned compared to a naive brute-force search. Our evaluation will cover multiple aspects of performance, such as construction time, query time, and the impact of incremental insertions of index efficiency. We will use multiple values of $k$ in the nearest neighbor search so that our results are relevant to multiple use-cases. Toward this end we will also use multiple datasets and distance metrics to further validate our findings. 

\subsection{Evaluation Procedure}

The approach used in most prior works to evaluate metric indexes is to create the index from all of the data, and then query each datapoint in the index search for the single nearest neighbor \cite{Izbicki2015}. For consistency we replicate this experiment style, but do not use every datapoint as a query point. This results in worst case $O(n^2)$ runtime for some of our tests, preventing us from comparing on our larger datasets. Since our interest is in if the index allows for faster queries, we can instead determined this the average pruning efficiency with extreme accuracy by using only small sample of query points. \todo[color=yellow]{fragment? -CNK Fixed? -Ed}
In tests using a sample of 1000 points for testing, versus using all data points, we found no difference in conclusions or results\footnote{The largest observed discrepancy was of 0.3 percentage points}. Thus we will use 1000 test points in all experiments. This will allow us to run any individual test in under a week, and evaluate the insertion-query performance in a more timely manner.

When using various datasets, if the dataset has a standard validation set, it will not be used. Instead points from the training set will be used for querying. This is done for constituency since not every dataset has a standard validation or testing set. 
Our experiments will be performed searching for the $k$ nearest neighbors with $k \in \{1, 5, 25, 100\}$. Evaluating for multiple values of $k$ is often ignored in most works, which focus on the $k=1$ case in their experiments \citep[e.g.][]{Izbicki2015,Cayton2012,Yianilos1993}, or will test on only a few small value of $k \leq 10$ \cite{Beygelzimer2006}. This is despite many applications, such as embeddings for visualization \cite{pmlr-v28-tarlow13,Maaten2008,VanderMaaten2014,Tang:2016:VLH:2872427.2883041}, using values of $k$ as large as 100. By testing a range of values for $k$ we can determine if one algorithm is uniformly better for all values of $k$, or if different algorithms have an advantage in one regime over the others. 

To evaluate the impact of incremental index construction on the quality of the final index, each index will be constructed in three different ways. Differences in performance between these three versions of the index will indicate the relative impact that incremental insertions have. 
\begin{enumerate}
\item Using the whole dataset and performing the classic batch construction method, by which we mean the original index construction process for each algorithm (referred to as batch construction)
\item Using half the dataset to construct an initial index using the classic batch method, and incrementally inserting the second half of the data (referred to as half-batch)
\item Constructing the entire dataset incrementally (referred to as incremental). 
\end{enumerate}
For these experiments, the Cover-tree is excluded --- as its original batch construction is already incremental (though does not support efficient queries between insertions). In our results we will expect the RBC algorithm to have minimal change in performance, due to the stochastic nature of representative selection. The expected performance impact of the VP-tree is unknown, though we would expect the tree to perform best in batch construction, second best when using half-batch construction, and worst when fully incremental. Results will consider both the number of distance computations when including and excluding distanced performed during index construction. We note that runtime of all methods and tests correlates directly with number of distance computations done for our code. Comparing distance computations is preferred so that we observe the true impact of pruning, rather than efficiency of micro optimizations, and is thus comparable to implementations written in other languages. 

We will also test the effectiveness of each method when interleaving queries and insertions. This will be evaluated in a manner analogous to common data structures, where we have different number of possible read (query) and write (insert) ratios.

\subsection{Data and Distances Used }

Now that we have reviewed how we will evaluate our methods, we will list the datasets and distance metrics used in such evaluations. A summary of which is presented in \autoref{tbl:datasets}. Datasets and distance metrics were selected to cover a wide range of data and metric types, include common baselines, and so that experiments would finish within a one-week execution window.   

\begin{table}[!htb]
\centering
\begin{tabular}{@{}lrc@{}}
\toprule
\multicolumn{1}{c}{Dataset} & \multicolumn{1}{c}{Samples} & Distance Metric \\ \midrule
MNIST                       & 60,000                      & Euclidean       \\
MNIST8m                     & 8,000,000                   & Euclidean       \\
Covtype                     & 581,012                     & Euclidean       \\
VxHeaven                    & 271,095                     & LZJD            \\
VirusShare5m                & 5,000,000                   & LZJD            \\
ILSVRC 2012 Validation      & 50,000                      & EMD             \\
IMDB Movie Titles           & 143,337                     & Levenshtein     \\
\bottomrule
\end{tabular}
\caption{Datasets used in experiments, including the number of points in each dataset and the distance metric used.}
\label{tbl:datasets}
\end{table}

Our first three datasets will all use the familiar euclidean distance\eqref{eq:euclidean}. The first of which is the well known MNIST dataset \cite{726791}, which is a commonly used benchmark for machine learning in general. Due to its small size we also include a larger version of the dataset, MNIST8m, which contains 8 million points produced by random transformations to the original dataset \cite{loosli-canu-bottou-2006}. We also evaluate the Forest Cover Type (Covtype) datasets \cite{Blackard1999}, which has historically been used for metric indexes. 
\begin{equation}\label{eq:euclidean}
d(x, y) = \lVert x-y \rVert 
\end{equation}

Finding nearest neighbors and similar examples is important for malware analysis \cite{Jang2011,Hu:2009:LMI:1653662.1653736}. The VxHeaven corpus has been widely used for research in malware analysis \cite{vxheaven}, and so we use it in our work for measuring the similarity of binaries. VxHeaven contains 271k binaries, but malware datasets are routinely reaching the hundreds of millions to billions of samples. For this reason we also select a random 5 million element set from the VirusShare corpus \cite{VirusShare}, which shares real malware with interested researchers. As the distance metric for these datasets, we will use the Lempel-Ziv Jaccard Distance (LZJD) \cite{raff_lzjd_2017}, which was designed for measuring binary similarity and is based upon the Jaccard distance. LZJD uses the Lempel-Ziv algorithm to break a byte sequence up into a set of sub-sequences, and then uses the Jaccard distance \eqref{eq:jaccard_sim} to measure the distance between these sets. Recent work has used LZJD for related tasks such as similarity digests for digital forensics, where prior tools could not be accelerated in the same manner since they lacked the distance metric properties \cite{raff_lzjd_digest}. 

\begin{equation}\label{eq:jaccard_sim}
d(A, B) = 1-\frac{|A \cap B |}{|A \cup B|}
\end{equation}

One of the metrics measured in the original Cover-tree paper was the a string edit distance \cite{Beygelzimer2006}. They compared to the dataset and methods used in \citet{Clarkson2002}, however the available data contains only 200 test strings. Instead we use the Levenshtein edit distance on IMDB movie titles \cite{Behm:2011:AAS:2004686.2005592}, which contains both longer strings and is three orders of magnitude larger. 

The simplified Cover-tree paper evaluated a larger range of distance metrics \cite{Izbicki2015}, including the Earth Mover's Distance (EMD) \cite{Rubner2000}. The EMD provides a distance measure between histograms, and was originally proposed for measuring the similarity of images. We follow the same procedure as \author{Izbicki2015} for using the "thresholded" EMD \cite{Pele-iccv2009}, except we use the RGB color space\footnote{Our software did not support the LabCIE color space previously used, and we did not notice any significant difference in results for other color spaces.}. We use the 2012 validation set of the ImageNet challenge \cite{ILSVRC15} for this distance metric, as it is the most computationally demanding metric of the ones we evaluate in this work. 

\section{Evaluation of Construction Time and Pruning Improvements} \label{sec:eval_nn_prune_improve}

We first evaluate the impact of our changes to each of the three algorithms. For RBC and VP-trees, we have made alterations that aim to improve the ability of these algorithms to avoid unnecessary distance computations at query time. For the Cover-tree, we have made a modification that will negatively impact its ability to perform pruning, but will make it viable for interleaved insertions and queries. We will evaluate the impact of our changes on construction time, query efficiency under normal construction, and the impact incremental construction has on the efficiency of the complete index. 

\subsection{Impact on Construction Time}

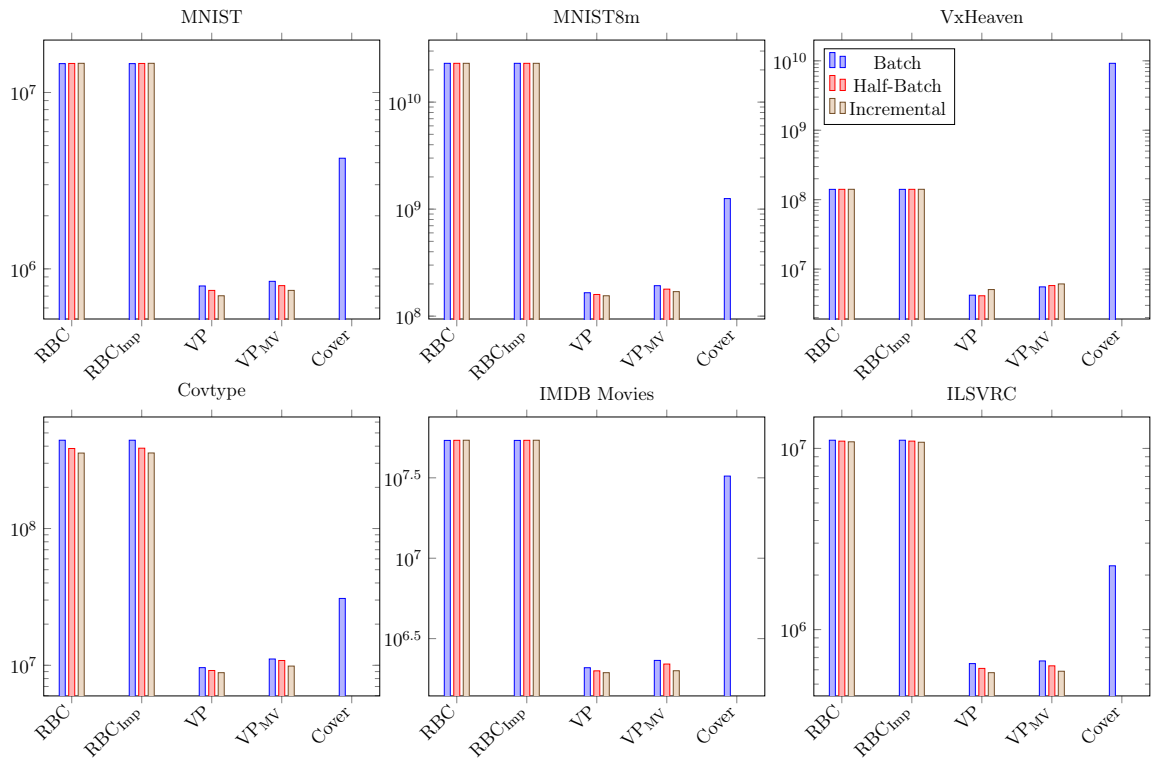
\begin{figure}[!htb]
\centering
\begin{adjustbox}{max size={\textwidth}{\textheight}}
  \begin{tikzpicture}
  \begin{groupplot}[
      group style={
        	group name=myplot, group size=3 by 2,
			vertical sep=2cm,
        },
        enlarge x limits=true,
      ]
    \centering
    \nextgroupplot[title=MNIST, %
                ybar,
                bar width=3.5,
                ymode = log,
                symbolic x coords={RBC_orig, RBC, VP, VPMV,
                    Cover, CoverB
                },
                xtick=data,
                xticklabels={RBC,
                    \rbci,
                    VP,
                    \vpmv,
                    Cover,
                    \coverb
                },
                x tick label style={rotate=45, anchor=east, align=right},
    	]
            \addplot table[x=algo,y=norm_dists,col sep=comma]{logs/MNIST_construction.csv};
            \addplot table[x=algo,y=half_dists,col sep=comma]{logs/MNIST_construction2.csv};
            \addplot table[x=algo,y=inc_dists,col sep=comma]{logs/MNIST_construction2.csv};
    
    \nextgroupplot[title=MNIST8m, %
                ybar,
                bar width=3.5,
                ymode = log,
                symbolic x coords={RBC_orig, RBC, VP, VPMV,
                    Cover, CoverB
                },
                xtick=data,
                xticklabels={RBC,
                    \rbci,
                    VP,
                    \vpmv,
                    Cover,
                    \coverb
                },
                x tick label style={rotate=45, anchor=east, align=right},
    	]
            \addplot table[x=algo,y=norm_dists,col sep=comma]{logs/MNIST8m_construction.csv};
            \addplot table[x=algo,y=half_dists,col sep=comma]{logs/MNIST8m_construction2.csv};
            \addplot table[x=algo,y=inc_dists,col sep=comma]{logs/MNIST8m_construction2.csv};
    
   \nextgroupplot[title=VxHeaven,
                ybar,
                bar width=3.5,
                ymode = log,
                symbolic x coords={RBC_orig, RBC, VP, VPMV,
                    Cover, CoverB
                },
                xtick=data,
                xticklabels={RBC,
                    \rbci,
                    VP,
                    \vpmv,
                    Cover,
                    \coverb
                },
                x tick label style={rotate=45, anchor=east, align=right},
                legend pos=north west,
            ]
            \addplot table[x=algo,y=norm_dists,col sep=comma]{logs/VxHeaven_construction.csv};
            \addplot table[x=algo,y=half_dists,col sep=comma]{logs/VxHeaven_construction2.csv};
            \addplot table[x=algo,y=inc_dists,col sep=comma]{logs/VxHeaven_construction2.csv};
            \legend{Batch, Half-Batch, Incremental}
    
  \nextgroupplot[title=Covtype,
                ybar,
                bar width=3.5,
                ymode = log,
                symbolic x coords={RBC_orig, RBC, VP, VPMV,
                    Cover, CoverB
                },
                xtick=data,
                xticklabels={RBC,
                    \rbci,
                    VP,
                    \vpmv,
                    Cover,
                    \coverb
                },
                x tick label style={rotate=45, anchor=east, align=right},
            ]
            \addplot table[x=algo,y=norm_dists,col sep=comma]{logs/Covtype_construction.csv};
            \addplot table[x=algo,y=half_dists,col sep=comma]{logs/Covtype_construction2.csv};
            \addplot table[x=algo,y=inc_dists,col sep=comma]{logs/Covtype_construction2.csv};
    
  \nextgroupplot[title=IMDB Movies,
                ybar,
                bar width=3.5,
                ymode = log,
                symbolic x coords={RBC_orig, RBC, VP, VPMV,
                    Cover, CoverB
                },
                xtick=data,
                xticklabels={RBC,
                    \rbci,
                    VP,
                    \vpmv,
                    Cover,
                    \coverb
                },
                x tick label style={rotate=45, anchor=east, align=right},
            ]
            \addplot table[x=algo,y=norm_dists,col sep=comma]{logs/IMDB_construction.csv};
            \addplot table[x=algo,y=half_dists,col sep=comma]{logs/IMDB_construction2.csv};
            \addplot table[x=algo,y=inc_dists,col sep=comma]{logs/IMDB_construction2.csv};
            
  \nextgroupplot[title=ILSVRC,
                ybar,
                bar width=3.5,
                ymode = log,
                symbolic x coords={RBC_orig, RBC, VP, VPMV,
                    Cover, CoverB
                },
                xtick=data,
                xticklabels={RBC,
                    \rbci,
                    VP,
                    \vpmv,
                    Cover,
                    \coverb
                },
                x tick label style={rotate=45, anchor=east, align=right},
            ]
            \addplot table[x=algo,y=norm_dists,col sep=comma]{logs/ImgNetVal_construction.csv};
            \addplot table[x=algo,y=half_dists,col sep=comma]{logs/ImgNetVal_construction2.csv};
            \addplot table[x=algo,y=inc_dists,col sep=comma]{logs/ImgNetVal_construction2.csv};
    
    \end{groupplot}

  \end{tikzpicture}
\end{adjustbox}
  \caption{Construction performance for each algorithm on each dataset. The y-axis represents the number of distance computations performed to build each index. Each algorithm is plotted three times, once using classic batch construction, half-batch, and incremental. The Cover-tree's construction algorithm is equivalent in all scenarios, so only one bar is shown.}
  \label{fig:construction_time_dists}
\end{figure}

To determine the impact of the incremental construction and our modifications, we will compare each algorithm in terms of the number of distance computations needed to construct the index. 
We will do this for all three construction options, batch, half-batch, and incremental, as discussed in \autoref{sec:data_method}. The time for only constructing the indices in these three ways are shown in \autoref{fig:construction_time_dists}. We note that there is no distinction between the Cover and \coverb construction times, and that the cover-tree is always incremental in construction. For this reason we only show one bar to represent Cover and \coverb across all three construction scenarios to avoid graph clutter.

Here we see the two performance characteristics observed. On datasets like MNIST, where we use the euclidean distance, RBC is the slowest to construct. This is expected, as it also has the highest complexity at $O(n \sqrt{n} )$ time. We also note that the RBC radius search is not as efficient at pruning, and fails to do so on most datasets. Only on datasets that are most accelerated, such as the Covtype dataset, does the RBC incremental construction avoid distance computations during construction. This  empirically supports the theoretical justification that we maintain the same construction time for the RBC algorithm, as discussed in \autoref{sec:rbc_inc_construction}. 

The second slowest to construct is the Cover-tree, followed by the VP-trees which is fastest. On the VxHeaven dataset, with the LZJD metric, the construction time performance of the Cover-tree degrades dramatically, using two orders of magnitude more distance computations than the RBC. We believe this performance degradation is an artifact of the expansion constant $c$ that occurs when using the LZJD metric. 
\todo[inline, color=yellow]{But the expansion constant isn't a property of LZJD, is it? -CNK No, in review i stated it was related to data and metric combined}
The VP tree has no construction time impact with $c$, and the RBC algorithm has a small $O(c^{3/2})$ dependency compared to the Cover-tree's $O(c^6)$ dependence. On the VirusShare5m dataset, the Cover-tree couldn't be constructed given over a month of compute time. We also note that the Cover-tree had degraded construction performance on the IMDB Movies dataset using the Levenshtein distance. These results present a potential weakness in the Cover-tree algorithm. 

Barring the performance behavior of the Cover-tree, both the RBC and VP-tree have more consistent performance on various datasets. We note of particular interest that the incremental construction procedure for the RBC results in almost no change in the number of distance computations needed to build the index\footnote{The same cannot be said for wall clock time, which is expected.}. The radius search is rarely able to do any pruning for the RBC algorithm, and so the brute force degrades to the same number of distance computations as the batch insertion. The Covtype dataset is the one for which each algorithm was able to do the most pruning, and thus has the most pronounced effect of this. 

The \vpmv variant of the VP-tree also matches the construction profile of the standard VP-tree on each dataset, with slightly increased or decreased computations depending on the dataset. This is to be expected, as the standard VP-tree always produces balanced splits during batch construction. The incremental construction can also cause lopsided splits for both the VP and \vpmv-tree, which results in a longer recurrence during construction, and thus increased construction time and distances. The \vpmv-tree may also encourage such lopsided splits, increasing the occurrence of this behavior. Simultaneously, the incremental construction requires fewer distance computations to determine early splits, and so can result in fewer overall computations if the splits happen to come out near balanced. The data and metric dependent properties will determine which impact is stronger for a given case. The impact of incremental construction on the VP-trees is also variable, and can increase or decrease construction time. In either direction, the change in VP construction time is minor relative to the costs for Cover-trees and the RBC algorithm. 

Overall we can draw the following conclusions about construction time efficiency. 1) that the VP-trees are fastest in all cases, and the proposed \vpmv variant has no detrimental impact. 2) the RBC algorithms are the most consistent, but often slowest, and that the \rbci has no detrimental impact. 3) the Cover-tree is not consistent in its performance relative to the other two algorithms, but when it works well, is in the middle of the road. 

\subsection{Impact on Batch Query Efficiency}

We now look at the impact of our changes to the three search procedures on querying the index, when the index is built in the standard batch manner. This isolates the change in performance to only our modifications of the three algorithms. 
Our goal here is to show that \rbci and \vpmv are improvements over the standard RBC and VP-tree methods. We also want to quantify the negative impact of using the looser bounds in \coverb that will allow for incremental insertion and querying, which is not easy
with the standard simplified Cover-tree due to its use of the \textit{maxdist} bound and potential restructuring on insertions \cite{Izbicki2015}. 

\begin{figure}[!htb]
\centering
\begin{adjustbox}{max size={\textwidth}{\textheight}}
  \begin{tikzpicture}
  \begin{groupplot}[
      group style={
        	group name=myplot, group size=3 by 2,
            vertical sep=1.5cm,
        },
        enlarge x limits=true,
      ]
    \centering
    \nextgroupplot[title=MNIST, legend to name=grouplegend,
    	xmode = log,
    ]
    \addplot +[mark=o, color=blue, discard if not={algo}{RBC}] table [y=distAllR, x=k,col sep=comma]{logs/MNIST_query.csv}; \label{plots:rbc}
    \addplot +[dashed, mark=diamond, mark options={scale=2,solid}, color=Turquoise, discard if not={algo}{RBCi}] table [y=distAllR, x=k,col sep=comma]{logs/MNIST_query.csv}; \label{plots:rbci}
    \addplot +[mark=o, color=red, discard if not={algo}{VP}] table [y=distAllR, x=k,col sep=comma]{logs/MNIST_query.csv}; \label{plots:vp}
    \addplot +[dashed, mark=diamond, mark options={scale=2,solid}, color=RubineRed, discard if not={algo}{VPMV}] table [y=distAllR, x=k,col sep=comma]{logs/MNIST_query.csv}; \label{plots:vpmv}
    \addplot +[mark=o, color=ForestGreen, discard if not={algo}{Cover}] table [y=distAllR, x=k,col sep=comma]{logs/MNIST_query.csv}; \label{plots:cover}
    \addplot +[dashed, mark=diamond, mark options={scale=2,solid}, color=LimeGreen, discard if not={algo}{CoverB}] table [y=distAllR, x=k,col sep=comma]{logs/MNIST_query.csv}; \label{plots:coverb}
    \addplot[dotted, mark=none, black, samples=2] coordinates {(1.0,1.0) (100,1.0)};
    
    \nextgroupplot[title=MNIST8m, legend to name=grouplegend,
        ymode = log,
    	xmode = log,
    ]
    \addplot +[mark=o, color=blue, discard if not={algo}{RBC}] table [y=distAllR, x=k,col sep=comma]{logs/MNIST8m_query.csv};
    \addplot +[dashed, mark=diamond, mark options={scale=2,solid}, color=Turquoise, discard if not={algo}{RBCi}] table [y=distAllR, x=k,col sep=comma]{logs/MNIST8m_query.csv}; 
    \addplot +[mark=o, color=red, discard if not={algo}{VP}] table [y=distAllR, x=k,col sep=comma]{logs/MNIST8m_query.csv}; 
    \addplot +[dashed, mark=diamond, mark options={scale=2,solid}, color=RubineRed, discard if not={algo}{VPMV}] table [y=distAllR, x=k,col sep=comma]{logs/MNIST8m_query.csv}; 
    \addplot +[mark=o, color=ForestGreen, discard if not={algo}{Cover}] table [y=distAllR, x=k,col sep=comma]{logs/MNIST8m_query.csv}; 
    \addplot +[dashed, mark=diamond, mark options={scale=2,solid}, color=LimeGreen, discard if not={algo}{CoverB}] table [y=distAllR, x=k,col sep=comma]{logs/MNIST8m_query.csv}; 
    \addplot[dotted, mark=none, black, samples=2] coordinates {(1.0,1.0) (100,1.0)};
    
   \nextgroupplot[title=VxHeaven,
      xmode = log,
      ]
    \addplot +[mark=o, color=blue, discard if not={algo}{RBC}] table [y=distAllR, x=k,col sep=comma]{logs/VxHeaven_query.csv};
    \addplot +[dashed, mark=diamond, mark options={scale=2,solid}, color=Turquoise, discard if not={algo}{RBCi}] table [y=distAllR, x=k,col sep=comma]{logs/VxHeaven_query.csv};
    \addplot +[mark=o, color=red, discard if not={algo}{VP}] table [y=distAllR, x=k,col sep=comma]{logs/VxHeaven_query.csv};
    \addplot +[dashed, mark=diamond, mark options={scale=2,solid}, color=RubineRed, discard if not={algo}{VPMV}] table [y=distAllR, x=k,col sep=comma]{logs/VxHeaven_query.csv};
    \addplot +[mark=o, color=ForestGreen, discard if not={algo}{Cover}] table [y=distAllR, x=k,col sep=comma]{logs/VxHeaven_query.csv};
    \addplot +[dashed, mark=diamond, mark options={scale=2,solid}, color=LimeGreen, discard if not={algo}{CoverB}] table [y=distAllR, x=k,col sep=comma]{logs/VxHeaven_query.csv};
    \addplot[dotted, mark=none, black, samples=2] coordinates {(1.0,1.0) (100,1.0)};

  \nextgroupplot[title=Covtype,
      ymode = log,
      xmode = log,
      xlabel={$k$},
      ]
    \addplot +[mark=o, color=blue, discard if not={algo}{RBC}] table [y=distAllR, x=k,col sep=comma]{logs/Covtype_query.csv};
    \addplot +[dashed, mark=diamond, mark options={scale=2,solid}, color=Turquoise, discard if not={algo}{RBCi}] table [y=distAllR, x=k,col sep=comma]{logs/Covtype_query.csv};
    \addplot +[mark=o, color=red, discard if not={algo}{VP}] table [y=distAllR, x=k,col sep=comma]{logs/Covtype_query.csv};
    \addplot +[dashed, mark=diamond, mark options={scale=2,solid}, color=RubineRed, discard if not={algo}{VPMV}] table [y=distAllR, x=k,col sep=comma]{logs/Covtype_query.csv};
    \addplot +[mark=o, color=ForestGreen, discard if not={algo}{Cover}] table [y=distAllR, x=k,col sep=comma]{logs/Covtype_query.csv};
    \addplot +[dashed, mark=diamond, mark options={scale=2,solid}, color=LimeGreen, discard if not={algo}{CoverB}] table [y=distAllR, x=k,col sep=comma]{logs/Covtype_query.csv};
    \addplot[dotted, mark=none, black, samples=2] coordinates {(1.0,1.0) (100,1.0)};

    \nextgroupplot[title=IMDB Movies,
      xmode = log,
      xlabel={$k$},
      ]
    \addplot +[mark=o, color=blue, discard if not={algo}{RBC}] table [y=distAllR, x=k,col sep=comma]{logs/IMDB_query.csv};
    \addplot +[dashed, mark=diamond, mark options={scale=2,solid}, color=Turquoise, discard if not={algo}{RBCi}] table [y=distAllR, x=k,col sep=comma]{logs/IMDB_query.csv};
    \addplot +[mark=o, color=red, discard if not={algo}{VP}] table [y=distAllR, x=k,col sep=comma]{logs/IMDB_query.csv};
    \addplot +[dashed, mark=diamond, mark options={scale=2,solid}, color=RubineRed, discard if not={algo}{VPMV}] table [y=distAllR, x=k,col sep=comma]{logs/IMDB_query.csv};
    \addplot +[mark=o, color=ForestGreen, discard if not={algo}{Cover}] table [y=distAllR, x=k,col sep=comma]{logs/IMDB_query.csv};
    \addplot +[dashed, mark=diamond, mark options={scale=2,solid}, color=LimeGreen, discard if not={algo}{CoverB}] table [y=distAllR, x=k,col sep=comma]{logs/IMDB_query.csv};
    \addplot[dotted, mark=none, black, samples=2] coordinates {(1.0,1.0) (100,1.0)};
    
    \nextgroupplot[title=ILSVRC,
      ymode = log,
      xmode = log,
      xlabel={$k$},
      ]
    \addplot +[mark=o, color=blue, discard if not={algo}{RBC}] table [y=distAllR, x=k,col sep=comma]{logs/ImgNetVal_query.csv};
    \addplot +[dashed, mark=diamond, mark options={scale=2,solid}, color=Turquoise, discard if not={algo}{RBCi}] table [y=distAllR, x=k,col sep=comma]{logs/ImgNetVal_query.csv};
    \addplot +[mark=o, color=red, discard if not={algo}{VP}] table [y=distAllR, x=k,col sep=comma]{logs/ImgNetVal_query.csv};
    \addplot +[dashed, mark=diamond, mark options={scale=2,solid}, color=RubineRed, discard if not={algo}{VPMV}] table [y=distAllR, x=k,col sep=comma]{logs/ImgNetVal_query.csv};
    \addplot +[mark=o, color=ForestGreen, discard if not={algo}{Cover}] table [y=distAllR, x=k,col sep=comma]{logs/ImgNetVal_query.csv};
    \addplot +[dashed, mark=diamond, mark options={scale=2,solid}, color=LimeGreen, discard if not={algo}{CoverB}] table [y=distAllR, x=k,col sep=comma]{logs/ImgNetVal_query.csv};
    \addplot[dotted, mark=none, black, samples=2] coordinates {(1.0,1.0) (100,1.0)};
    
    \end{groupplot}

  \path (myplot c1r1.outer north west)%
            -- node[anchor=south,rotate=90] {Fraction of Distance Computations Needed}%
            (myplot c1r2.outer south west)%
      ;
  \path (myplot c1r1.north west|-current bounding box.north)--
        coordinate(legendpos)
        (myplot c3r1.north east|-current bounding box.north);
  \matrix[
      matrix of nodes,
      anchor=south,
      draw,
      inner sep=0.2em,
      draw
    ]at([yshift=1ex]legendpos)
    {
      \ref{plots:rbc}& RBC &[5pt]
      \ref{plots:rbci}& \rbci &[5pt]
      \ref{plots:vp}& VP &[5pt]
      \ref{plots:vpmv}& \vpmv &[5pt]
      \ref{plots:cover}& Cover &[5pt]
      \ref{plots:coverb}& \coverb \\};

  \end{tikzpicture}
\end{adjustbox}
\caption{Number of distance computations needed as a function of the desired number of neighbors $k$. The y-axis is the ratio of distance computations compared to a brute-force search (shown at 1.0 as a dotted black line).  }
\label{fig:query_results_batch}
\end{figure}
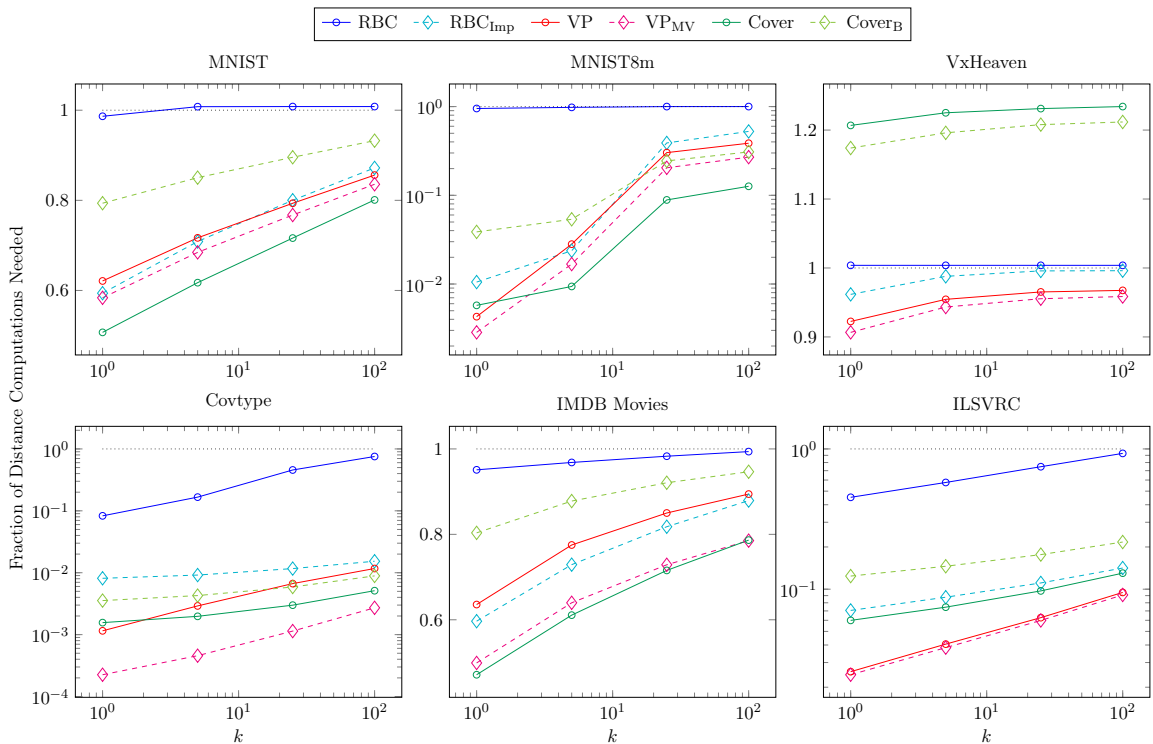

Considering only batch construction, we can see the query efficiency of these methods in \autoref{fig:query_results_batch}, where we look at the fraction of distance computations needed compared to a brute-force search. This figure factors in the distance computations needed during construction time, so the query efficiency is with respect to the whole process. 

We remind the reader that this plot is construed from a random sample of 1000 randomly selected query points, and then scaled to have the same weight as if all test points were used. That is to say, if a corpus as $n$ data points, we compute the average number of distance computations from a sample of 1000 points. The total number of distance computations is then treated as this average times $n$. This closely mimics the same results that would have been achieved by using all $n$ points as queries, but keeps runtime manageable given our compute resources. In extended testing on corpora where it is feasible to compute this for all $n$ points, just 100 samples reliably estimated the ratio to two significant figures, so our 1000 point estimates should allow us to reach the same conclusions with confidence. 

One can see that for the RBC and VP-tree algorithms, our enchantments to the search procedure are effective. For the RBC algorithm in particular, more distance computations were done than the brute force search in most cases, but \rbci  dramatically improves the competitiveness of the approach. This comes at a loss of compute efficiency when using the euclidean metric, which is where the RBC obtains its original speed improvements. But our work is looking at the general efficiencies of the RBC for arbitrary distance metrics, which may not have the same efficiency advantages when answering queries in batches. In this respect the pruning improvements of \rbci are dramatic and important if the RBC algorithm is to be used. 

The \vpmv reduces the number of computations needed compared to the standard VP-tree in all cases. The amount of improvement varies by dataset, ranging from almost no improvement, to nearly an order of magnitude less distance computations for the Covtype dataset. Given these results our choice to produce unbalanced splits during construction is empirically validated. 

As expected, the \coverb variant of the simplified Cover-tree had a detrimental impact on efficiency, as it is relaxing the bound to the same one used in the original Cover-tree work\cite{Beygelzimer2006}. Among all tests, the \coverb-tree required 1.6 to 6.7 times as many distance computations as the standard Cover-tree, with the exact values given in \autoref{tbl:coverb_impact} for all tested values of $k$. The few distance computations avoided for determining the tighter bound clearly make up for a considerable portion of the simplified Cover-tree's improved performance.

\begin{table}[!htb]
\centering
\caption{For each dataset, the this table shows the multipler on the number of distance computations \coverb had to perform compared to a normal Cover-tree.}
\label{tbl:coverb_impact}
\begin{tabular}{@{}rcccccc@{}}
\toprule
\multicolumn{1}{c}{}  & \multicolumn{6}{c}{Dataset}                               \\
\cmidrule(lr){2-7}
\multicolumn{1}{c}{k} & MNIST & MNIST8m & ILSVRC      & Covtype & IMDB & VxHeaven \\ 
\midrule
1                     & 1.57  & 6.73    & 2.07        & 2.27    & 1.70 & 0.97     \\
5                     & 1.38  & 5.71    & 1.96        & 2.16    & 1.44 & 0.98     \\
25                    & 1.25  & 2.75    & 1.81        & 1.97    & 1.29 & 0.98     \\
100                   & 1.16  & 2.44    & 1.67        & 1.73    & 1.20 & 0.98     \\
\bottomrule
\end{tabular}
\end{table}

While the Cover-tree was the most efficient at avoiding distance computations on the MNIST dataset, the Cover-tree is the worst performer by far on the VxHeaven dataset. The increased construction time results in the Cover-tree performing 20\% more distance computations than would be necessary with the brute force approach. 
We also see an interesting artifact that more distance computations were done on VxHeaven when using the tighter maxdist bound than the looser \coverb approach. This comes from the extra computations needed to obtain the maxdist bound in the first place, and indicates that more distances computations are being done to obtain that bound then are saved in more efficient pruning.

\begin{figure}[ht]
\centering
\begin{tikzpicture}[scale=0.95]
\begin{axis}[
      xmode = log,
      xlabel={$k$},
      ]
    \addplot +[mark=o, color=blue, discard if not={algo}{RBC}] table [y=distAllR, x=k,col sep=comma]{logs/VirusShare5m_query.csv};
    \addplot +[dashed, mark=diamond, mark options={scale=2,solid}, color=Turquoise, discard if not={algo}{RBCi}] table [y=distAllR, x=k,col sep=comma]{logs/VirusShare5m_query.csv};
    \addplot +[mark=o, color=red, discard if not={algo}{VP}] table [y=distAllR, x=k,col sep=comma]{logs/VirusShare5m_query.csv};
    \addplot +[dashed, mark=diamond, mark options={scale=2,solid}, color=RubineRed, discard if not={algo}{VPMV}] table [y=distAllR, x=k,col sep=comma]{logs/VirusShare5m_query.csv};
    \addplot[dotted, mark=none, black, samples=2] coordinates {(1.0,1.0) (100,1.0)};
\end{axis}
\end{tikzpicture}
\caption{Query performance on the VirusShare5m dataset. } 
\label{fig:virusshare5m_query}
\end{figure}
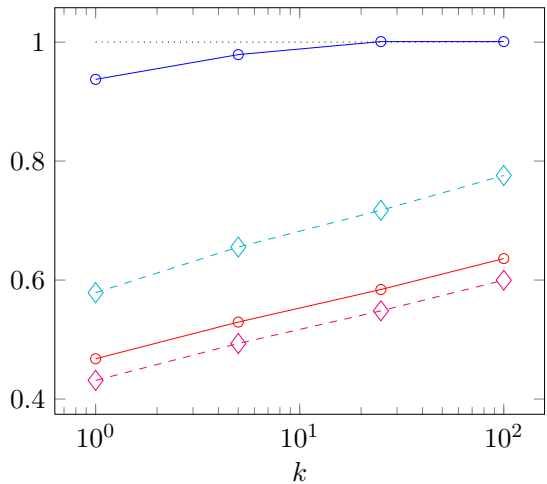

We also note that the VxHeaven dataset, using the LZJD distance, had the worst query performance amongst all datasets, with LZJD barely managing to avoid 5\% of the distance computations compared to a brute-force search. By testing this on the larger VirusShare5m dataset, as seen in \autoref{fig:virusshare5m_query}, we can see that increasing the corpus size does lead to pruning efficiencies. While the Cover-tree couldn't be built on this corpus, both the RBC and VP algorithms are able to perform reasonably well. The \vpmv did best, avoiding between 57\% and 40\% of the distance computations a brute-force search would require. 

Viewing these results as a whole, we would have to recommend the \vpmv algorithm as the best choice in terms of query efficiency. In all cases it either prunes the most distances for all values of $k$, or is a close second to the Cover-tree (which has an extreme failure case with LZJD). 

\subsection{Impact of Incremental Construction on Query Efficiency} \label{sec:impact_inc_con_query_eff}

For the last part of this section, we examine the impact on query pruning based on how the index was constructed. That is to say, does half-batch or incremental construction of the index negatively impact the ability to prune distance computations, and if so, by how much? Such evaluation will be shown for only the more efficient \rbci and \vpmv algorithms that we will further evaluate in \autoref{sec:eval_inc_const}. We do not consider the Cover-tree variants in this portion. As noted in \autoref{sec:cover_tree}, the Cover-tree's construction is already incremental. Thus these indexes will be equivalent when given the same insertion ordering. The only change in Cover-tree efficiency would be from random variance caused by changes in insertion order.

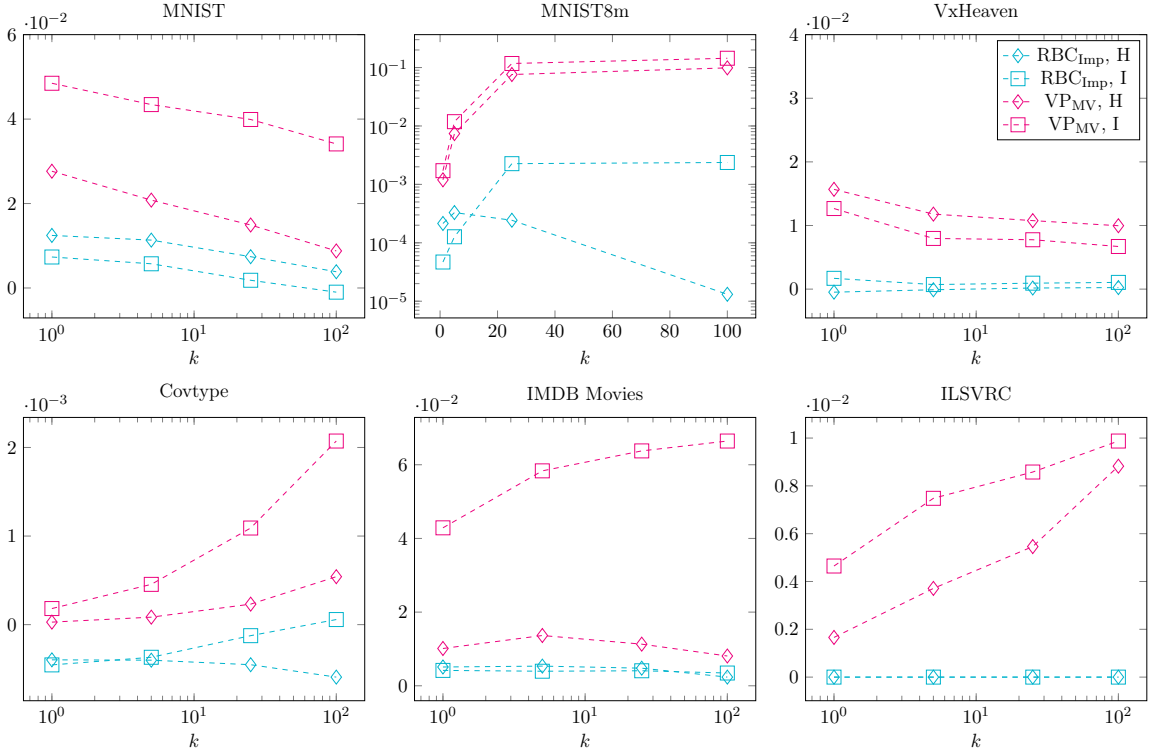
\begin{figure}[!htb]
\centering
\begin{adjustbox}{max size={\textwidth}{\textheight}}
  \begin{tikzpicture}
  \begin{groupplot}[
      group style={
        	group name=myplot, group size=3 by 2,
			vertical sep=2cm,
        },
        enlarge x limits=true,
      ]
    \centering
    \nextgroupplot[title=MNIST, legend to name=grouplegend,
	  ymax=0.06,
      xmode = log,
      xlabel={$k$},
      ]
      \addplot +[dashed, mark=diamond, mark options={scale=2,solid}, color=Turquoise, discard if not={algo}{RBCi}] table [y expr=\thisrowno{13}-\thisrowno{7}, x=k,col sep=comma]{logs/MNIST_query.csv};
      \addplot +[dashed, mark=square, mark options={scale=2,solid}, color=Turquoise, discard if not={algo}{RBCi}] table [y expr=\thisrowno{19}-\thisrowno{7}, x=k,col sep=comma]{logs/MNIST_query.csv};

      \addplot +[dashed, mark=diamond, mark options={scale=2,solid}, color=RubineRed, discard if not={algo}{VPMV}] table [y expr=\thisrowno{13}-\thisrowno{7}, x=k,col sep=comma]{logs/MNIST_query.csv};
      \addplot +[dashed, mark=square, mark options={scale=2,solid}, color=RubineRed, discard if not={algo}{VPMV}] table [y expr=\thisrowno{19}-\thisrowno{7}, x=k,col sep=comma]{logs/MNIST_query.csv};

    \nextgroupplot[title=MNIST8m, legend to name=grouplegend,
      ymode = log,
      xlabel={$k$},
      ]
      \addplot +[dashed, mark=diamond, mark options={scale=2,solid}, color=Turquoise, discard if not={algo}{RBCi}] table [y expr=\thisrowno{7}-\thisrowno{13}, x=k,col sep=comma]{logs/MNIST8m_query.csv};
      \addplot +[dashed, mark=square, mark options={scale=2,solid}, color=Turquoise, discard if not={algo}{RBCi}] table [y expr=\thisrowno{7}-\thisrowno{19}, x=k,col sep=comma]{logs/MNIST8m_query.csv};

      \addplot +[dashed, mark=diamond, mark options={scale=2,solid}, color=RubineRed, discard if not={algo}{VPMV}] table [y expr=\thisrowno{13}-\thisrowno{7}, x=k,col sep=comma]{logs/MNIST8m_query.csv};
      \addplot +[dashed, mark=square, mark options={scale=2,solid}, color=RubineRed, discard if not={algo}{VPMV}] table [y expr=\thisrowno{19}-\thisrowno{7}, x=k,col sep=comma]{logs/MNIST8m_query.csv};

   \nextgroupplot[title=VxHeaven,
      ymax=0.04,
      xmode = log,
      xlabel={$k$},
      ]
        \addplot +[dashed, mark=diamond, mark options={scale=2,solid}, color=Turquoise, discard if not={algo}{RBCi}] table [y expr=\thisrowno{13}-\thisrowno{7}, x=k,col sep=comma]{logs/VxHeaven_query.csv};
        \addlegendentry{\rbci, H}
        \addplot +[dashed, mark=square, mark options={scale=2,solid}, color=Turquoise, discard if not={algo}{RBCi}] table [y expr=\thisrowno{19}-\thisrowno{7}, x=k,col sep=comma]{logs/VxHeaven_query.csv};
        \addlegendentry{\rbci, I}

        \addplot +[dashed, mark=diamond, mark options={scale=2,solid}, color=RubineRed, discard if not={algo}{VPMV}] table [y expr=\thisrowno{13}-\thisrowno{7}, x=k,col sep=comma]{logs/VxHeaven_query.csv};
        \addlegendentry{\vpmv, H}
        \addplot +[dashed, mark=square, mark options={scale=2,solid}, color=RubineRed, discard if not={algo}{VPMV}] table [y expr=\thisrowno{19}-\thisrowno{7}, x=k,col sep=comma]{logs/VxHeaven_query.csv};
        \addlegendentry{\vpmv, I}
    
  \nextgroupplot[title=Covtype,
      xmode = log,
      xlabel={$k$},
      ]
        \addplot +[dashed, mark=diamond, mark options={scale=2,solid}, color=Turquoise, discard if not={algo}{RBCi}] table [y expr=\thisrowno{13}-\thisrowno{7}, x=k,col sep=comma]{logs/Covtype_query.csv};
        \addplot +[dashed, mark=square, mark options={scale=2,solid}, color=Turquoise, discard if not={algo}{RBCi}] table [y expr=\thisrowno{19}-\thisrowno{7}, x=k,col sep=comma]{logs/Covtype_query.csv};

        \addplot +[dashed, mark=diamond, mark options={scale=2,solid}, color=RubineRed, discard if not={algo}{VPMV}] table [y expr=\thisrowno{13}-\thisrowno{7}, x=k,col sep=comma]{logs/Covtype_query.csv};
        \addplot +[dashed, mark=square, mark options={scale=2,solid}, color=RubineRed, discard if not={algo}{VPMV}] table [y expr=\thisrowno{19}-\thisrowno{7}, x=k,col sep=comma]{logs/Covtype_query.csv};

  \nextgroupplot[title=IMDB Movies,
      xmode = log,
      xlabel={$k$},
      ]
        \addplot +[dashed, mark=diamond, mark options={scale=2,solid}, color=Turquoise, discard if not={algo}{RBCi}] table [y expr=\thisrowno{13}-\thisrowno{7}, x=k,col sep=comma]{logs/IMDB_query.csv};
        \addplot +[dashed, mark=square, mark options={scale=2,solid}, color=Turquoise, discard if not={algo}{RBCi}] table [y expr=\thisrowno{19}-\thisrowno{7}, x=k,col sep=comma]{logs/IMDB_query.csv};

        \addplot +[dashed, mark=diamond, mark options={scale=2,solid}, color=RubineRed, discard if not={algo}{VPMV}] table [y expr=\thisrowno{13}-\thisrowno{7}, x=k,col sep=comma]{logs/IMDB_query.csv};
        \addplot +[dashed, mark=square, mark options={scale=2,solid}, color=RubineRed, discard if not={algo}{VPMV}] table [y expr=\thisrowno{19}-\thisrowno{7}, x=k,col sep=comma]{logs/IMDB_query.csv};

\nextgroupplot[title=ILSVRC,
      xmode = log,
      xlabel={$k$},
      ]
        \addplot +[dashed, mark=diamond, mark options={scale=2,solid}, color=Turquoise, discard if not={algo}{RBCi}] table [y expr=\thisrowno{13}-\thisrowno{7}, x=k,col sep=comma]{logs/ImgNetVal_query.csv};
        \addplot +[dashed, mark=square, mark options={scale=2,solid}, color=Turquoise, discard if not={algo}{RBCi}] table [y expr=\thisrowno{19}-\thisrowno{7}, x=k,col sep=comma]{logs/ImgNetVal_query.csv};

        \addplot +[dashed, mark=diamond, mark options={scale=2,solid}, color=RubineRed, discard if not={algo}{VPMV}] table [y expr=\thisrowno{13}-\thisrowno{7}, x=k,col sep=comma]{logs/ImgNetVal_query.csv};
        \addplot +[dashed, mark=square, mark options={scale=2,solid}, color=RubineRed, discard if not={algo}{VPMV}] table [y expr=\thisrowno{19}-\thisrowno{7}, x=k,col sep=comma]{logs/ImgNetVal_query.csv};

    \end{groupplot}

  \end{tikzpicture}
\end{adjustbox}
\caption{Difference in the number of distance computations needed as a function of the desired number of neighbors $k$. The y-axis is the difference in the ratio of distance computations compared to a brute-force search. We note that the scale on the y-axis is different for various figures, and the small scale indicates that incremental construction has little impact on query efficiency.  }
\label{fig:insertion_query_impact}
\end{figure}

The difference between the ratio of distance computations done for Half-Batch (H) and Incremental (I) index construction is shown in \autoref{fig:insertion_query_impact}. That is to say, if $r_H = \frac{\text{Distance Computations with Half-Batch}}{\text{Distance Computations Brute Force}}$, and $r_B$ has the same definition but for the Batch construction, then the y-axis of the figure shows $r_B-r_H$. This is also plot for the difference between incremental construction, i.e., $r_B-r_I$. When this value is near zero, it means that both the Batch and Half-Batch or Incremental construction approaches have avoided a similar number of distance computations. 

We remind the reader that Half-Batch is where the dataset is constructed using the standard batch construction approach for the first $n/2$ data-points, and the remaining $n/2$ are inserted incrementally. Incremental construction builds the index from empty to full using only the incremental insertions. 

Positive values indicate an increase in the number of distance queries needed. Negative values indicate a reduction in the number of distance queries needed, and are generally an indication of problem variance. That is to say, when the difference in ratios can go negative, it's because the natural variance (caused by insertion order randomness) is greater than the impact of the incremental construction. Such scenarios would generally be considered favorable, as it would indicate that our modifications have no particular positive or negative impact. 

We first note a general pattern in that the difference in query efficiency can go up or down with changes in the desired number of neighbors $k$. This will be an artifact of both the dataset and distance metric used, and highlights the importance of testing metric structures over a large range of $k$. Testing over a wide range of $k$ has not been historically done in previous works, usually performing only the $1-nn$ search. 

In our results we can see that the RBC algorithm  performs best in these tests. The \rbci approach's pruning ability is minimally impacted by changes in construction for all datasets and values of $k$. The largest increase is on MNIST for $k=1$, where the Half-Batch insertion scenario increases from 59.4\% to 60.6\%, an increase of only 1.2 percentage points. It  makes sense that the \rbci approach would have a consistent minimal degradation in query efficiency, as the structure of the RBC is coarse, and our incremental insertion strategy closely matches the behavior of the batch creation strategy. 

The \vpmv-tree does not perform as well as the \rbci, and we can see that incremental construction always has a more larger, but still small, impact on its performance for all datasets. The only case where this exceeds a two percentage point difference is on the MNIST8m dataset, where a $\approx7.6\%$ point gap occurs for incremental and half-batch construction. The larger impact on the \vpmv's performance is understandable given that our insertion procedure does not have the same information available for choosing splits, which may cause sub-optimal choices. 

Our expectation would be that the \vpmv's performance would degrade more when using incremental (I) insertion rather than half-batch (H), as the half-batch insertion will get to use more datapoints to estimate the split point for nodes higher up in the tree. Our results generally support this hypothesis, with \vpmv (I) causing  more distance queries to be performance than the (H) case. However, for MNIST8m, VxHeaven, and ILSVRC, the performance gap is not that large across the tested values of $k$. This suggests that the loosened bounds during insertion may also be an issue impacting the efficiency after insertions. One possible way to reduce this impact would be to add multiple vantage points dynamically during insertion, to avoid impacting the existing low/high bounds of the VP-tree. Such Multi-Vantage-Point (MVP) trees have been explored previously\cite{Bozkaya1999} in a batch construction context. We leave research in exploiting such extensions to to future work. 

Regarding the impact on query efficiency given incremental insertions, we can confidently state that the RBC approach is well poised to this part of the problem, with almost no negative impact to efficiency. The VP-tree does not fair quite as well, but is still more efficient than the \rbci algorithm in all of these cases after construction from only incremental insertions. 

Overall, we can draw some immediate conclusions with respect to our proposed changes to Cover-trees, VP-trees, and the RBC index. First, that VP-trees in general strike a strong balance between construction time cost and query time efficiency across many datasets with differing metrics. For both the RBC and VP tree, we can improve their query time efficiency across the board. These improvements come with minimal cost, and so we can consider them exclusively in \autoref{sec:eval_inc_const} where we look at incremental insertions and querying. We also observe that the Cover-tree is significantly degraded at insertion/construction time by when using the LZJD distance. 

\section{Evaluation of Incremental Insertion-Query Efficiency} \label{sec:eval_inc_const}

At this point we have shown that \rbci and \vpmv are improvements over the original RBC and VP-tree algorithms in terms of query efficiency, with no significant impact on the construction time. We have also shown that the indexes constructed by them are still effective are pruning distance computations, which encourages their use. We can now evaluate their overall effectiveness when we interleave insertions and queries in a single system. 

In this section we now consider the case of evaluating each index from the context of incremental insertion and querying. Contrasting with the standard scenario, where we build an index and immediately query it (usually for k-nearest neighbor classification, or some similar purpose), we will be building an index and evaluating the number of distance computations performed after construction. This scenario corresponds to many realistic use cases, where a large training set is deployed for use, and new data added to the index over time. 

Given a dataset with $n$ items in it, our evaluation procedure will consider $r$ queries (or "reads") and $w$ insertions (or "writes") to the index. The naive case, where we perform brute force search, there is no cost to writing to the index, only when we perform a query. 
This brute force approach also represents our baseline for the maximum number of distance computations needed to answer the queries. 

Similar to data structures for storing and accessing data and concurrency tools, we may also explore differing ratios of reads to writes. In our experiments we evaluated insert/query ratios from 100:1 to 1:100. In all cases, we found that the most challenging scenario was when we had 100 insertions for each query. This is not surprising, as all of our data structures have a non-zero cost for insertions, and in the case of RBC and Cover-trees, can be quite significant. Thus, below we will only present results for the case where we have 100 insertions for each query, and our tests will limited to 1000 insertions due to runtime constraints\footnote{We allowed a maximum of one week runtime for tests to complete in this scenario.}. We construct each initial index on half of the data points, using the batch construction method. 

\begin{figure}[!htb]
\centering
\begin{adjustbox}{max size={\textwidth}{\textheight}}
  \begin{tikzpicture}
  \begin{groupplot}[
      group style={
        	group name=myplot, group size=3 by 2,
			vertical sep=2cm,
        },
        enlarge x limits=true,
      ]
    \centering
    \nextgroupplot[title=MNIST,
	  xmode = log,
      xlabel={$k$},
      legend columns=1,
      legend style={/tikz/every even column/.append style={column sep=0.5cm}},
      legend pos=south east, 
      ]
      \addplot +[dashed, mark=diamond, mark options={scale=2,solid}, color=Turquoise, discard if nottwo={algo}{RBCi}{rw}{0.01}] table [y=allDistR, x=k,col sep=comma]{logs/MNIST_insertion.csv};
\addlegendentry{\rbci}

      \addplot +[dashed, mark=diamond, mark options={scale=2,solid}, color=RubineRed, discard if nottwo={algo}{VPMV}{rw}{0.01}] table [y=allDistR, x=k,col sep=comma]{logs/MNIST_insertion.csv};
      \addlegendentry{\vpmv}

      \addplot +[dashed, mark=diamond, mark options={scale=2,solid}, color=LimeGreen, discard if nottwo={algo}{CoverB}{rw}{0.01}] table [y=allDistR, x=k,col sep=comma]{logs/MNIST_insertion.csv};
      \addlegendentry{\coverb}
      
      \addplot +[dashed, mark=star, mark options={scale=2,solid}, color=LimeGreen, discard if nottwo={algo}{CoverI}{rw}{0.01}] table [y=allDistR, x=k,col sep=comma]{logs/MNIST_insertion.csv};
      \addlegendentry{\coveri}

      \addplot[dotted, mark=none, black, samples=2] coordinates {(1.0,1.0) (100,1.0)};
    
    \nextgroupplot[title=MNIST8m,
      ymode = log,
	  xmode = log,
      xlabel={$k$},
      ]
      \addplot +[dashed, mark=diamond, mark options={scale=2,solid}, color=Turquoise, discard if nottwo={algo}{RBCi}{rw}{0.01}] table [y=allDistR, x=k,col sep=comma]{logs/MNIST8m_insertion.csv};

      \addplot +[dashed, mark=diamond, mark options={scale=2,solid}, color=RubineRed, discard if nottwo={algo}{VPMV}{rw}{0.01}] table [y=allDistR, x=k,col sep=comma]{logs/MNIST8m_insertion.csv};

      \addplot +[dashed, mark=diamond, mark options={scale=2,solid}, color=LimeGreen, discard if nottwo={algo}{CoverB}{rw}{0.01}] table [y=allDistR, x=k,col sep=comma]{logs/MNIST8m_insertion.csv};
      
      \addplot +[dashed, mark=star, mark options={scale=2,solid}, color=LimeGreen, discard if nottwo={algo}{CoverI}{rw}{0.01}] table [y=allDistR, x=k,col sep=comma]{logs/MNIST8m_insertion.csv};

      \addplot[dotted, mark=none, black, samples=2] coordinates {(1.0,1.0) (100,1.0)};
    
   \nextgroupplot[title=VxHeaven,
      xmode = log,
      xlabel={$k$},
      ]
      \addplot +[dashed, mark=diamond, mark options={scale=2,solid}, color=Turquoise, discard if nottwo={algo}{RBCi}{rw}{0.01}] table [y=allDistR, x=k,col sep=comma]{logs/VxHeaven_insertion.csv};

      \addplot +[dashed, mark=diamond, mark options={scale=2,solid}, color=RubineRed, discard if nottwo={algo}{VPMV}{rw}{0.01}] table [y=allDistR, x=k,col sep=comma]{logs/VxHeaven_insertion.csv};

      \addplot +[dashed, mark=diamond, mark options={scale=2,solid}, color=LimeGreen, discard if nottwo={algo}{CoverB}{rw}{0.01}] table [y=allDistR, x=k,col sep=comma]{logs/VxHeaven_insertion.csv};

		\addplot +[dashed, mark=star, mark options={scale=2,solid}, color=LimeGreen, discard if nottwo={algo}{CoverI}{rw}{0.01}] table [y=allDistR, x=k,col sep=comma]{logs/VxHeaven_insertion.csv};

      \addplot[dotted, mark=none, black, samples=2] coordinates {(1.0,1.0) (100,1.0)};
    
  \nextgroupplot[title=Covtype,
      xmode = log,
      ymode = log,
      xlabel={$k$},
      legend columns=1,
      legend style={/tikz/every even column/.append style={column sep=0.5cm}},
      legend pos=south east, 
      ]
      \addplot +[dashed, mark=diamond, mark options={scale=2,solid}, color=Turquoise, discard if nottwo={algo}{RBCi}{rw}{0.01}] table [y=allDistR, x=k,col sep=comma]{logs/Covtype_insertion.csv};

      \addplot +[dashed, mark=diamond, mark options={scale=2,solid}, color=RubineRed, discard if nottwo={algo}{VPMV}{rw}{0.01}] table [y=allDistR, x=k,col sep=comma]{logs/Covtype_insertion.csv};

      \addplot +[dashed, mark=diamond, mark options={scale=2,solid}, color=LimeGreen, discard if nottwo={algo}{CoverB}{rw}{0.01}] table [y=allDistR, x=k,col sep=comma]{logs/Covtype_insertion.csv};

		\addplot +[dashed, mark=star, mark options={scale=2,solid}, color=LimeGreen, discard if nottwo={algo}{CoverI}{rw}{0.01}] table [y=allDistR, x=k,col sep=comma]{logs/Covtype_insertion.csv};

  \nextgroupplot[title=IMDB Movies,
      xmode = log,
      xlabel={$k$},
      legend columns=1,
      legend style={/tikz/every even column/.append style={column sep=0.5cm}},
      legend pos=south east, 
      ]
      \addplot +[dashed, mark=diamond, mark options={scale=2,solid}, color=Turquoise, discard if nottwo={algo}{RBCi}{rw}{0.01}] table [y=allDistR, x=k,col sep=comma]{logs/IMDB_insertion.csv};

      \addplot +[dashed, mark=diamond, mark options={scale=2,solid}, color=RubineRed, discard if nottwo={algo}{VPMV}{rw}{0.01}] table [y=allDistR, x=k,col sep=comma]{logs/IMDB_insertion.csv};

      \addplot +[dashed, mark=diamond, mark options={scale=2,solid}, color=LimeGreen, discard if nottwo={algo}{CoverB}{rw}{0.01}] table [y=allDistR, x=k,col sep=comma]{logs/IMDB_insertion.csv};

		\addplot +[dashed, mark=star, mark options={scale=2,solid}, color=LimeGreen, discard if nottwo={algo}{CoverI}{rw}{0.01}] table [y=allDistR, x=k,col sep=comma]{logs/IMDB_insertion.csv};

      \addplot[dotted, mark=none, black, samples=2] coordinates {(1.0,1.0) (100,1.0)};
    
  \nextgroupplot[title=ILSVRC,
      xmode = log,
      xlabel={$k$},
      legend columns=1,
      legend style={/tikz/every even column/.append style={column sep=0.5cm}},
      legend pos=south east, 
      ]
      \addplot +[dashed, mark=diamond, mark options={scale=2,solid}, color=Turquoise, discard if nottwo={algo}{RBCi}{rw}{0.01}] table [y=allDistR, x=k,col sep=comma]{logs/ImgNetVal_insertion.csv};

      \addplot +[dashed, mark=diamond, mark options={scale=2,solid}, color=RubineRed, discard if nottwo={algo}{VPMV}{rw}{0.01}] table [y=allDistR, x=k,col sep=comma]{logs/ImgNetVal_insertion.csv};

      \addplot +[dashed, mark=diamond, mark options={scale=2,solid}, color=LimeGreen, discard if nottwo={algo}{CoverB}{rw}{0.01}] table [y=allDistR, x=k,col sep=comma]{logs/ImgNetVal_insertion.csv};

		\addplot +[dashed, mark=star, mark options={scale=2,solid}, color=LimeGreen, discard if nottwo={algo}{CoverI}{rw}{0.01}] table [y=allDistR, x=k,col sep=comma]{logs/ImgNetVal_insertion.csv};

    \end{groupplot}

  \end{tikzpicture}
\end{adjustbox}
\caption{Fraction of distance computations needed (relative to naive approach) in incremental scenario, with 100 insertions for every query. Does not include \textit{initial} construction costs, only subsequent insertion costs.  }
\label{fig:incremental_query}
\end{figure}
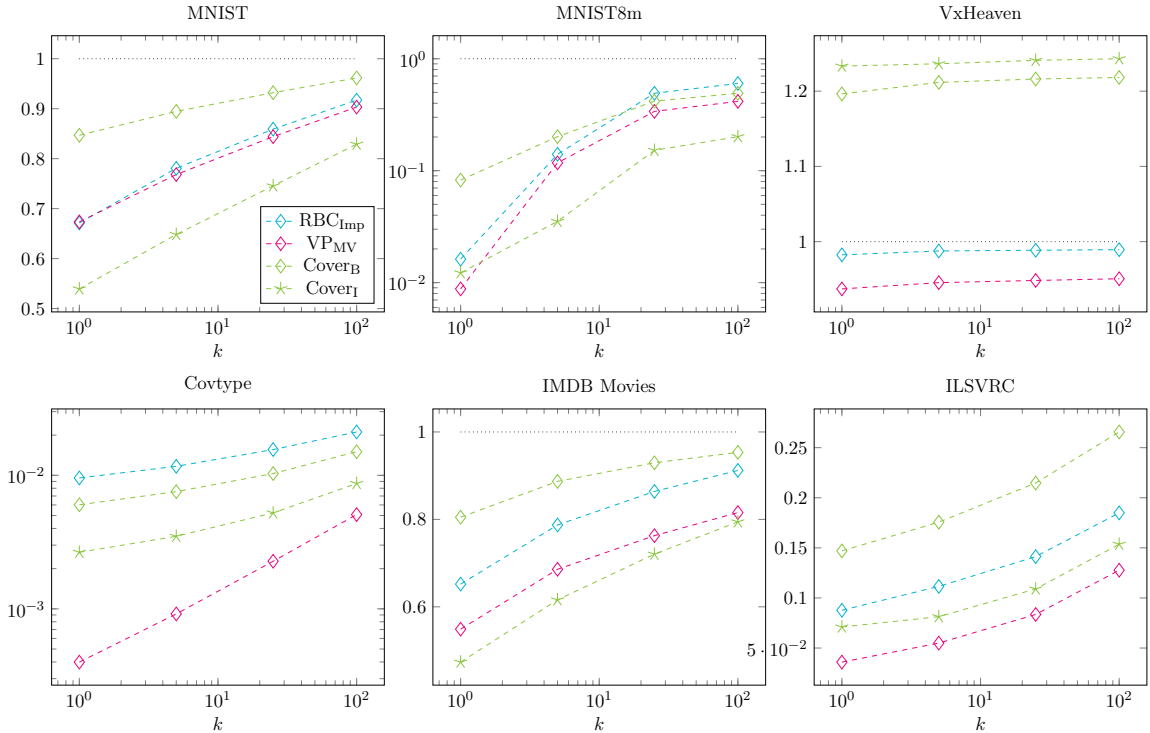

For the Cover-tree, only \coverb produces reasonable insertion/query performance, as the \textit{maxdist} bound can't be maintained when re-balancing occurs. Using the original loose bound causes a considerable reducing in efficiency at query time. By recording the multiplicative difference between the tighter bound Cover-tree and the original looser bound in \coverb in \autoref{tbl:coverb_impact}, we can plot the performance of the \textit{ideal} Cover-tree as a function of \coverb. This gives us a measure of what the best possible performance of the Cover-tree would be in this scenario, as it ignores all overheads in any potential scheme for selectively updating the Cover-tree bound as items are inserted that would cause re-balancing. We will indicate this ideal Cover-tree as \coveri.

The results of our methods are presented in \autoref{fig:incremental_query}. Amongst the \rbci, \vpmv, and \coverb algorithms, the \vpmv dominates all other approaches. It successfully avoids the most total distance computations to answer nearest neighbor queries for all values of $k$ on all datasets. This is not surprising given the cumulative results of \autoref{sec:eval_nn_prune_improve}, which found the \vpmv to require the fewest distance computations during construction time and was always either the most efficient at avoiding distance computations, or nearly behind the Cover-tree approach. 

If we had an ability to obtain the \textit{maxdist} bound for free, we can also see that the \coveri approach is still not very competitive with the \vpmv-tree. While \coveri does have better performance than \vpmv on some datasets, it often trails behind on the Covtype by nearly an order of magnitude. Especially when we consider the failure of the Cover-trees to perform with the LZJD distance on VxHeaven and VirusShare5m. This variability in performance makes the Cover-tree less desirable to use for arbitrary distance metrics. 

While the \vpmv appear to be the best overall fit to our task, we note that our \rbci also makes a strong showing despite the $O(\sqrt{n})$ complexity target instead of $O(\log(n))$. \rbci consistently performs better than random guessing, which can't be said for the Cover-tree. On the more difficult datasets, it is often not far behind the \vpmv-tree in performance, though it is an order of magnitude less efficient on the Covtype and ILSVRC datasets. The biggest weakness of the RBC  approach is that the incremental insertions will have an amortized cost, with the insertion time increasing dramatically every $\sqrt{n}$ insertions to expand the representative set. If the number of insertions is known to be bounded, this may be an avoidable cost -- thus increasing the RBC's practicality. We note as well that in the case of datasets stored in a distributed index across multiple server's, the RBC's coarse structure may allow for more efficient parallelization. This may be an important factor in future work when we consider datasets larger than what can be stored on a single machine. 

\subsection{Discussion}

While we have modified three algorithms for our scenario of incremental querying and insertion, we note that there is a further unexplored area for improvement in the "Read write" ratio. In our case it was most challenging for all algorithms to handle more "Writes" per "read", as each insertion required multiple distance computations and the insertions did not dramatically change the performance at query time. This is in part because we have modified existing algorithms to support this scenario, and so the performance interleaving insertions and queries closely follows the performance when we evaluate query by including the construction cost, as we did in \autoref{sec:eval_nn_prune_improve}.

Of the algorithms we have tested the \vpmv performs best with the lowest construction time, and is almost always the fastest at query time. This is also in the context of evaluation in a single-threaded scenario. When we consider a multi-threaded scenario, the \vpmv can utilize multiple threads for index construction using the batch-construction approach. However, insertion of a single data-point cannot easily be parallelized. The Cover-tree also has this challenge. 

Our \rbci approach presents a potential advantage over both of these algorithms when we consider the multi-thread or distributed scenario. As a consequence of how the RBC algorithm achieves its $O(\sqrt{n})$ insertion and query time, we can readily parallelize line 1 of \autoref{algo:rbc_insert} on up to $\sqrt{p}$ processors, requiring only a reduce operation to determine which processor had the closest representative. It may then be more practical than the \vpmv approach for extremely large indexes if sufficient compute resources are available. The downside to the RBC algorithm comes when the representative set must be increased, requiring more work and presenting a insertion cost that will periodically spike. This could be remedied by amortizing the cost of increasing the representative set across the preceding insertions, but we leave this to future work as we must consider the real-world efficiency of an implementation to determine how practical a solution it would be. 

In future work we hope to develop new algorithms that are specifically designed for incremental insertion and querying. We note two potential high level strategies in which one may develop methods that perform better for read and write heavy use-cases. We consider these beyond the scope of our current work, which looks at modifying existing algorithms, but may be fruitful inspiration for specialized methods. 

\subsubsection{Write \& Insert Heavy}

When we have multiple datapoints inserted before each query, it may become possible to use the index itself to accelerate the insertion process. Say that there will be a set of $Z$ points inserted into the index at a time. We can cluster the members of $Z$ by their density/closeness, and insert each cluster together as a group. One option may be to find the medoid of the group and its radius, which can then be used as a proxy point that represents the group as a whole. One could then insert the sub-groups into the index with a reduced number of distance computations if the triangle inequality can be used to determine that all members of the group belong in the same region of the index. The group may then be dissolved as such macro level pruning becomes impossible, or reduced into smaller sub-groups to continue the process. The dual-tree query approach \cite{Curtin:2014:DFE:2913413.2913416}, at a high level, presents a similar strategy for efficiently answering multiple queries at a time. 

\subsubsection{Read \& Query Heavy}

Another scenario is that insertions into the index will be \todo[color=yellow]{huh? -CNK Fixed? -ER}relatively rare, compared to the amount of nearest neighbor queries given to the index. In this case it may be desired to have the query process itself build and restructure the tree. This notion is in a similar spirit to splay trees and the union-find algorithm \cite{Tarjan:1984:WAS:62.2160,Tarjan:1975:EGB:321879.321884,Hopcroft1973}. Insertions to the dataset would be placed in a convenient location, and their first distances computed when a new query is given. Say that $x_i$ was a previously inserted point. Once we have a new query $x_q$, the distance to the query is obtained for the $x_i$ and for $x_q$'s nearest neighbors. If $d(x_i, x_q) \approx c \cdot d(x_q, x^{(k)})$, where $x^{(k)}$ is $x_q$'s $k$'th nearest neighbor and $c$ is some constant, we can then infer that $x_i$ should be placed in a similar location in the index. As multiple insertions are performed, we can use these distances with respect to the query to determine which points are related and should be kept close in the index. 

\section{Conclusions and Future Work} \label{sec:conclusions}

We have now evaluated and improved three different algorithms, Cover Trees, Vantage-Point Trees, Random Ball Covers, for the use case of incremental insertions and querying. We have significantly improved the query efficiency of the later two with our new \rbci and \vpmv variants, and introduced schemes to incrementally add to these collections.  Evaluation of all these methods was done with a number of datasets with varying sizes using four different distance metrics. In doing so, we can conclude that the \vpmv tree provides the best overall performance for our task. It requires the fewest distance computations during construction, is consistently one of the fastest at query time, and this balance produces the best overall results when interleaving insertions and queries. 

While already successful, the \vpmv tree still has room for improvement. It has the highest degradation to performance from insertions, which could perhaps be remedied by a smarter update algorithm or the use of multiple vantage points. While the \coverb algorithm could be improved by obtaining a better alternative bound than \textit{maxdist}, it appears obtaining a computational cheaper version \textit{maxdist} bound itself is not sufficient to remedy the performance gap when using the LZJD distance.

\bibliography{Mendeley}

\begin{thebibliography}{54}
\providecommand{\natexlab}[1]{#1}
\providecommand{\url}[1]{\texttt{#1}}
\expandafter\ifx\csname urlstyle\endcsname\relax
  \providecommand{\doi}[1]{doi: #1}\else
  \providecommand{\doi}{doi: \begingroup \urlstyle{rm}\Url}\fi

\bibitem[vxh()]{vxheaven}
{VX Heaven}.
\newblock URL \url{https://vxheaven.org/}.

\bibitem[Behm et~al.(2011)Behm, Li, and Carey]{Behm:2011:AAS:2004686.2005592}
A.~Behm, C.~Li, and M.~J. Carey.
\newblock {Answering Approximate String Queries on Large Data Sets Using
  External Memory}.
\newblock In \emph{Proceedings of the 2011 IEEE 27th International Conference
  on Data Engineering}, ICDE '11, pages 888--899, Washington, DC, USA, 2011.
  IEEE Computer Society.
\newblock ISBN 978-1-4244-8959-6.
\newblock \doi{10.1109/ICDE.2011.5767856}.
\newblock URL \url{http://dx.doi.org/10.1109/ICDE.2011.5767856}.

\bibitem[Bentley(1975)]{Bentley:1975:MBS:361002.361007}
J.~L. Bentley.
\newblock {Multidimensional Binary Search Trees Used for Associative
  Searching}.
\newblock \emph{Commun. ACM}, 18\penalty0 (9):\penalty0 509--517, 9 1975.
\newblock ISSN 0001-0782.
\newblock \doi{10.1145/361002.361007}.
\newblock URL \url{http://doi.acm.org/10.1145/361002.361007}.

\bibitem[Beygelzimer et~al.(2006)Beygelzimer, Kakade, and
  Langford]{Beygelzimer2006}
A.~Beygelzimer, S.~Kakade, and J.~Langford.
\newblock {Cover trees for nearest neighbor}.
\newblock In \emph{International Conference on Machine Learning}, pages
  97--104, New York, 2006. ACM.
\newblock URL
  \url{http://www.cs.princeton.edu/courses/archive/spr05/cos598E/bib/covertree.pdf}.

\bibitem[Bi{\c{c}}ici and Yuret(2007)]{Bicici2007}
E.~Bi{\c{c}}ici and D.~Yuret.
\newblock {Locally scaled density based clustering}.
\newblock In B.~Beliczynski, A.~Dzielinski, M.~Iwanowski, and B.~Ribeiro,
  editors, \emph{Adaptive and Natural Computing Algorithms}, page 739–748,
  Warsaw, Poland, 2007. Springer-Verlag.
\newblock URL \url{http://www.springerlink.com/index/0116171485446868.pdf}.

\bibitem[Blackard and Dean(1999)]{Blackard1999}
J.~A. Blackard and D.~J. Dean.
\newblock {Comparative accuracies of artificial neural networks and
  discriminant analysis in predicting forest cover types from cartographic
  variables}.
\newblock \emph{Computers and Electronics in Agriculture}, 24\penalty0
  (3):\penalty0 131--151, 1999.
\newblock ISSN 01681699.
\newblock \doi{10.1016/S0168-1699(99)00046-0}.
\newblock URL
  \url{http://www.sciencedirect.com/science/article/pii/S0168169999000460}.

\bibitem[Bozkaya and Ozsoyoglu(1999)]{Bozkaya1999}
T.~Bozkaya and M.~Ozsoyoglu.
\newblock {Indexing large metric spaces for similarity search queries}.
\newblock \emph{ACM Transactions on Database Systems (TODS)}, 24\penalty0
  (3):\penalty0 361--404, 1999.
\newblock URL \url{http://dl.acm.org/citation.cfm?id=328959}.

\bibitem[Breiman et~al.(1984)Breiman, Friedman, Stone, and Olshen]{Breiman1984}
L.~Breiman, J.~Friedman, C.~J. Stone, and R.~Olshen.
\newblock \emph{{Classification and Regression Trees}}.
\newblock CRC press, 1984.

\bibitem[Campello et~al.(2013)Campello, Moulavi, and Sander]{hdbscan}
R.~J. G.~B. Campello, D.~Moulavi, and J.~Sander.
\newblock {Density-Based Clustering Based on Hierarchical Density Estimates}.
\newblock In J.~Pei, V.~Tseng, L.~Cao, H.~Motoda, and G.~Xu, editors,
  \emph{Advances in Knowledge Discovery and Data Mining}, pages 160--172.
  Springer Berlin Heidelberg, 2013.
\newblock ISBN 978-3-642-37455-5.
\newblock \doi{10.1007/978-3-642-37456-2{\_}14}.
\newblock URL \url{http://link.springer.com/10.1007/978-3-642-37456-2_14}.

\bibitem[Cayton(2012)]{Cayton2012}
L.~Cayton.
\newblock {Accelerating Nearest Neighbor Search on Manycore Systems}.
\newblock \emph{IEEE 26th International Parallel and Distributed Processing
  Symposium}, pages 402--413, 5 2012.
\newblock \doi{10.1109/IPDPS.2012.45}.
\newblock URL
  \url{http://ieeexplore.ieee.org/lpdocs/epic03/wrapper.htm?arnumber=6267877}.

\bibitem[Chan et~al.(1983)Chan, Golub, and LeVeque]{Chan1983}
T.~F. Chan, G.~H. Golub, and R.~J. LeVeque.
\newblock {Algorithms for Computing the Sample Variance: Analysis and
  Recommendations}.
\newblock \emph{The American Statistician}, 37\penalty0 (3):\penalty0 242, 8
  1983.
\newblock ISSN 00031305.
\newblock \doi{10.2307/2683386}.
\newblock URL \url{http://www.jstor.org/stable/2683386?origin=crossref}.

\bibitem[Clarkson(2002)]{Clarkson2002}
K.~L. Clarkson.
\newblock {Nearest neighbor searching in metric spaces: Experimental Results
  for sb(S)}.
\newblock Technical report, Bell Laboratories, Lucent Technologies, New Jersey,
  2002.

\bibitem[Curtin and Ram(2014)]{Curtin:2014:DFE:2913413.2913416}
R.~R. Curtin and P.~Ram.
\newblock {Dual-tree Fast Exact Max-kernel Search}.
\newblock \emph{Statistical Analysis and Data Mining}, 7\penalty0 (4):\penalty0
  229--253, 8 2014.
\newblock ISSN 1932-1864.
\newblock \doi{10.1002/sam.11218}.
\newblock URL \url{http://dx.doi.org/10.1002/sam.11218}.

\bibitem[Finkel and Bentley(1974)]{Finkel:1974:QTD:2697709.2697865}
R.~A. Finkel and J.~L. Bentley.
\newblock {Quad Trees a Data Structure for Retrieval on Composite Keys}.
\newblock \emph{Acta Informatica}, 4\penalty0 (1):\penalty0 1--9, 3 1974.
\newblock ISSN 0001-5903.
\newblock \doi{10.1007/BF00288933}.
\newblock URL \url{http://dx.doi.org/10.1007/BF00288933}.

\bibitem[Fukunage and Narendra(1975)]{Fukunage1975}
K.~Fukunage and P.~Narendra.
\newblock {A Branch and Bound Algorithm for Computing k-Nearest Neighbors}.
\newblock \emph{IEEE Transactions on Computers}, C-24\penalty0 (7):\penalty0
  750--753, 1975.
\newblock ISSN 0018-9340.
\newblock \doi{10.1109/T-C.1975.224297}.
\newblock URL
  \url{http://ieeexplore.ieee.org/xpl/freeabs_all.jsp?arnumber=1672890}.

\bibitem[Gieseke et~al.(2014)Gieseke, Heinermann, Oancea, and
  Igel]{Gieseke:2014:BKT:3044805.3044826}
F.~Gieseke, J.~Heinermann, C.~Oancea, and C.~Igel.
\newblock {Buffer K-d Trees: Processing Massive Nearest Neighbor Queries on
  GPUs}.
\newblock In \emph{Proceedings of the 31st International Conference on
  International Conference on Machine Learning - Volume 32}, ICML'14, pages
  I--172--I--180. JMLR.org, 2014.
\newblock URL \url{http://dl.acm.org/citation.cfm?id=3044805.3044826}.

\bibitem[Gove et~al.(2014)Gove, Saxe, Gold, Long, and
  Bergamo]{Gove:2014:SSV:2671491.2671496}
R.~Gove, J.~Saxe, S.~Gold, A.~Long, and G.~Bergamo.
\newblock {SEEM: A Scalable Visualization for Comparing Multiple Large Sets of
  Attributes for Malware Analysis}.
\newblock In \emph{Proceedings of the Eleventh Workshop on Visualization for
  Cyber Security}, VizSec '14, pages 72--79, New York, NY, USA, 2014. ACM.
\newblock ISBN 978-1-4503-2826-5.
\newblock \doi{10.1145/2671491.2671496}.
\newblock URL \url{http://doi.acm.org/10.1145/2671491.2671496}.

\bibitem[Guttman(1984)]{Guttman1984}
A.~Guttman.
\newblock {R-trees: a dynamic index structure for spatial searching}.
\newblock In \emph{1984 ACM SIGMOD international conference on Management of
  data}, pages 47--57, New York, NY, 1984. ACM.
\newblock ISBN 0897911288.
\newblock URL \url{http://dl.acm.org/citation.cfm?id=602266}.

\bibitem[Hopcroft and Ullman(1973)]{Hopcroft1973}
J.~E. Hopcroft and J.~D. Ullman.
\newblock {Set Merging Algorithms}.
\newblock \emph{SIAM Journal on Computing}, 2\penalty0 (4):\penalty0 294--303,
  12 1973.
\newblock ISSN 0097-5397.
\newblock \doi{10.1137/0202024}.
\newblock URL \url{http://epubs.siam.org/doi/10.1137/0202024}.

\bibitem[Hu et~al.(2009)Hu, Chiueh, and Shin]{Hu:2009:LMI:1653662.1653736}
X.~Hu, T.-c. Chiueh, and K.~G. Shin.
\newblock {Large-scale Malware Indexing Using Function-call Graphs}.
\newblock In \emph{Proceedings of the 16th ACM Conference on Computer and
  Communications Security}, CCS '09, pages 611--620, New York, NY, USA, 2009.
  ACM.
\newblock ISBN 978-1-60558-894-0.
\newblock \doi{10.1145/1653662.1653736}.
\newblock URL \url{http://doi.acm.org/10.1145/1653662.1653736}.

\bibitem[Hu et~al.(2013)Hu, Shin, Bhatkar, and Griffin]{178990}
X.~Hu, K.~G. Shin, S.~Bhatkar, and K.~Griffin.
\newblock {MutantX-S: Scalable Malware Clustering Based on Static Features}.
\newblock In \emph{Presented as part of the 2013 USENIX Annual Technical
  Conference (USENIX ATC 13)}, pages 187--198, San Jose, CA, 2013. USENIX.
\newblock ISBN 978-1-931971-01-0.
\newblock URL
  \url{https://www.usenix.org/conference/atc13/technical-sessions/presentation/hu}.

\bibitem[Izbicki and Shelton(2015)]{Izbicki2015}
M.~Izbicki and C.~R. Shelton.
\newblock {Faster Cover Trees}.
\newblock In \emph{Proceedings of the Thirty-Second International Conference on
  Machine Learning}, volume~37, 2015.

\bibitem[Jang et~al.(2011)Jang, Brumley, and Venkataraman]{Jang2011}
J.~Jang, D.~Brumley, and S.~Venkataraman.
\newblock {BitShred: Feature Hashing Malware for Scalable Triage and Semantic
  Analysis}.
\newblock In \emph{Proceedings of the 18th ACM conference on Computer and
  communications security - CCS}, pages 309--320, New York, New York, USA,
  2011. ACM Press.
\newblock ISBN 9781450309486.
\newblock \doi{10.1145/2046707.2046742}.
\newblock URL \url{http://dl.acm.org/citation.cfm?doid=2046707.2046742}.

\bibitem[Kamel and Faloutsos(1994)]{Kamel:1994:HRI:645920.673001}
I.~Kamel and C.~Faloutsos.
\newblock {Hilbert R-tree: An Improved R-tree Using Fractals}.
\newblock In \emph{Proceedings of the 20th International Conference on Very
  Large Data Bases}, VLDB '94, pages 500--509, San Francisco, CA, USA, 1994.
  Morgan Kaufmann Publishers Inc.
\newblock ISBN 1-55860-153-8.
\newblock URL \url{http://dl.acm.org/citation.cfm?id=645920.673001}.

\bibitem[Kanungo et~al.(2002)Kanungo, Mount, Netanyahu, Piatko, Silverman, and
  Wu]{Kanungo2002}
T.~Kanungo, D.~Mount, N.~Netanyahu, C.~Piatko, R.~Silverman, and A.~Wu.
\newblock {An efficient k-means clustering algorithm: analysis and
  implementation}.
\newblock \emph{IEEE Transactions on Pattern Analysis and Machine
  Intelligence}, 24\penalty0 (7):\penalty0 881--892, 2002.
\newblock ISSN 0162-8828.
\newblock \doi{10.1109/TPAMI.2002.1017616}.

\bibitem[Karger and Ruhl(2002)]{Karger:2002:FNN:509907.510013}
D.~R. Karger and M.~Ruhl.
\newblock {Finding Nearest Neighbors in Growth-restricted Metrics}.
\newblock In \emph{Proceedings of the Thiry-fourth Annual ACM Symposium on
  Theory of Computing}, STOC '02, pages 741--750, New York, NY, USA, 2002. ACM.
\newblock ISBN 1-58113-495-9.
\newblock \doi{10.1145/509907.510013}.
\newblock URL \url{http://doi.acm.org/10.1145/509907.510013}.

\bibitem[Kim et~al.(2013)Kim, Kim, and Nam]{Kim2013}
J.~Kim, S.-G. Kim, and B.~Nam.
\newblock {Parallel multi-dimensional range query processing with R-trees on
  GPU}.
\newblock \emph{Journal of Parallel and Distributed Computing}, 73\penalty0
  (8):\penalty0 1195--1207, 2013.
\newblock ISSN 07437315.
\newblock \doi{10.1016/j.jpdc.2013.03.015}.
\newblock URL
  \url{http://www.sciencedirect.com/science/article/pii/S0743731513000592}.

\bibitem[Lecun et~al.(1998)Lecun, Bottou, Bengio, and Haffner]{726791}
Y.~Lecun, L.~Bottou, Y.~Bengio, and P.~Haffner.
\newblock {Gradient-based learning applied to document recognition}.
\newblock \emph{Proceedings of the IEEE}, 86\penalty0 (11):\penalty0
  2278--2324, 11 1998.
\newblock ISSN 0018-9219.
\newblock \doi{10.1109/5.726791}.

\bibitem[Li et~al.(2017)Li, Roundy, Gates, and Vorobeychik]{Li2017}
B.~Li, K.~Roundy, C.~Gates, and Y.~Vorobeychik.
\newblock {Large-Scale Identification of Malicious Singleton Files}.
\newblock In \emph{7TH ACM Conference on Data and Application Security and
  Privacy}, 2017.

\bibitem[Li and Malik(2016)]{Li2016}
K.~Li and J.~Malik.
\newblock {Fast k-Nearest Neighbour Search via Dynamic Continuous Indexing}.
\newblock In \emph{Proceedings of The 33rd International Conference on Machine
  Learning}, pages 671--679, 2016.
\newblock URL \url{http://arxiv.org/abs/1512.00442}.

\bibitem[Li and Amenta(2015)]{Li:2015:BKN:2977215.2977244}
S.~Li and N.~Amenta.
\newblock {Brute-Force k-Nearest Neighbors Search on the GPU}.
\newblock In \emph{Proceedings of the 8th International Conference on
  Similarity Search and Applications - Volume 9371}, SISAP 2015, pages
  259--270, New York, NY, USA, 2015. Springer-Verlag New York, Inc.
\newblock ISBN 978-3-319-25086-1.
\newblock \doi{10.1007/978-3-319-25087-8{\_}25}.
\newblock URL \url{http://dx.doi.org/10.1007/978-3-319-25087-8_25}.

\bibitem[Lloyd(1982)]{Lloyd1982}
S.~P. Lloyd.
\newblock {Least squares quantization in PCM}.
\newblock \emph{IEEE Transactions on Information Theory}, 28\penalty0
  (2):\penalty0 129--137, 3 1982.
\newblock ISSN 0018-9448.
\newblock \doi{10.1109/TIT.1982.1056489}.
\newblock URL
  \url{http://ieeexplore.ieee.org/lpdocs/epic03/wrapper.htm?arnumber=1056489}.

\bibitem[Loosli et~al.(2007)Loosli, Canu, and Bottou]{loosli-canu-bottou-2006}
G.~Loosli, S.~Canu, and L.~Bottou.
\newblock {Training Invariant Support Vector Machines using Selective
  Sampling}.
\newblock In L.~Bottou, O.~Chapelle, D.~DeCoste, and J.~Weston, editors,
  \emph{Large Scale Kernel Machines}, pages 301--320. MIT Press, Cambridge,
  MA., 2007.
\newblock URL \url{http://leon.bottou.org/papers/loosli-canu-bottou-2006}.

\bibitem[Maaten and Hinton(2008)]{Maaten2008}
L.~V.~D. Maaten and G.~Hinton.
\newblock {Visualizing Data using t-SNE}.
\newblock \emph{Journal of Machine Learning Research}, 9:\penalty0 2579--2605,
  2008.

\bibitem[Mohaisen and Alrawi(2013)]{Mohaisen:2013:UZA:2487788.2488056}
A.~Mohaisen and O.~Alrawi.
\newblock {Unveiling Zeus: Automated Classification of Malware Samples}.
\newblock In \emph{Proceedings of the 22Nd International Conference on World
  Wide Web}, WWW '13 Companion, pages 829--832, New York, NY, USA, 2013. ACM.
\newblock ISBN 978-1-4503-2038-2.
\newblock \doi{10.1145/2487788.2488056}.
\newblock URL \url{http://doi.acm.org/10.1145/2487788.2488056}.

\bibitem[Narayan et~al.(2015)Narayan, Punjani, and Abbeel]{Narayan2015}
K.~Narayan, A.~Punjani, and P.~Abbeel.
\newblock {Alpha-Beta Divergences Discover Micro and Macro Structures in Data}.
\newblock In \emph{Proceedings of The 32nd International Conference on Machine
  Learning}, pages 796--804, 2015.

\bibitem[Omohundro(1989)]{Omohundro1989}
S.~M. Omohundro.
\newblock {Five Balltree Construction Algorithms}.
\newblock Technical report, International Computer Science Institute, 1989.

\bibitem[Pele and Werman(2009)]{Pele-iccv2009}
O.~Pele and M.~Werman.
\newblock {Fast and robust Earth Mover's Distances}.
\newblock In \emph{2009 IEEE 12th International Conference on Computer Vision},
  pages 460--467. IEEE, 9 2009.
\newblock ISBN 978-1-4244-4420-5.
\newblock \doi{10.1109/ICCV.2009.5459199}.
\newblock URL
  \url{http://ieeexplore.ieee.org/lpdocs/epic03/wrapper.htm?arnumber=5459199}.

\bibitem[Raff and Nicholas(2017{\natexlab{a}})]{raff_lzjd_2017}
E.~Raff and C.~Nicholas.
\newblock {An Alternative to NCD for Large Sequences, Lempel-Ziv Jaccard
  Distance}.
\newblock In \emph{Proceedings of the 23rd ACM SIGKDD International Conference
  on Knowledge Discovery and Data Mining - KDD '17}, pages 1007--1015, New
  York, New York, USA, 2017{\natexlab{a}}. ACM Press.
\newblock ISBN 9781450348874.
\newblock \doi{10.1145/3097983.3098111}.
\newblock URL \url{http://dl.acm.org/citation.cfm?doid=3097983.3098111}.

\bibitem[Raff and Nicholas(2017{\natexlab{b}})]{raff_lzjd_digest}
E.~Raff and C.~K. Nicholas.
\newblock {Lempel-Ziv Jaccard Distance, an Effective Alternative to Ssdeep and
  Sdhash}.
\newblock \emph{arXiv preprint arXiv:1708.03346}, 8 2017{\natexlab{b}}.
\newblock URL \url{https://arxiv.org/abs/1708.03346}.

\bibitem[Roberts(2011)]{VirusShare}
J.-M. Roberts.
\newblock {Virus Share}, 2011.
\newblock URL \url{https://virusshare.com/}.

\bibitem[Rubner et~al.(2000)Rubner, Tomasi, and Guibas]{Rubner2000}
Y.~Rubner, C.~Tomasi, and L.~J. Guibas.
\newblock {The Earth Mover's Distance as a Metric for Image Retrieval}.
\newblock \emph{International Journal of Computer Vision}, 40\penalty0
  (2):\penalty0 99--121, 2000.
\newblock ISSN 09205691.
\newblock \doi{10.1023/A:1026543900054}.
\newblock URL \url{http://link.springer.com/10.1023/A:1026543900054}.

\bibitem[Russakovsky et~al.(2015)Russakovsky, Deng, Su, Krause, Satheesh, Ma,
  Huang, Karpathy, Khosla, Bernstein, Berg, and Fei-Fei]{ILSVRC15}
O.~Russakovsky, J.~Deng, H.~Su, J.~Krause, S.~Satheesh, S.~Ma, Z.~Huang,
  A.~Karpathy, A.~Khosla, M.~Bernstein, A.~C. Berg, and L.~Fei-Fei.
\newblock {ImageNet Large Scale Visual Recognition Challenge}.
\newblock \emph{International Journal of Computer Vision (IJCV)}, 115\penalty0
  (3):\penalty0 211--252, 2015.
\newblock \doi{10.1007/s11263-015-0816-y}.

\bibitem[Spafford(2014)]{tagkey2014iv}
E.~C. Spafford.
\newblock {Is Anti-virus Really Dead?}
\newblock \emph{Computers {\&} Security}, 44:\penalty0 iv, 2014.
\newblock ISSN 0167-4048.
\newblock \doi{http://dx.doi.org/10.1016/S0167-4048(14)00082-0}.
\newblock URL
  \url{http://www.sciencedirect.com/science/article/pii/S0167404814000820}.

\bibitem[Tang et~al.(2016)Tang, Liu, Zhang, and
  Mei]{Tang:2016:VLH:2872427.2883041}
J.~Tang, J.~Liu, M.~Zhang, and Q.~Mei.
\newblock {Visualizing Large-scale and High-dimensional Data}.
\newblock In \emph{Proceedings of the 25th International Conference on World
  Wide Web}, WWW '16, pages 287--297, Republic and Canton of Geneva,
  Switzerland, 2016. International World Wide Web Conferences Steering
  Committee.
\newblock ISBN 978-1-4503-4143-1.
\newblock \doi{10.1145/2872427.2883041}.
\newblock URL \url{http://dx.doi.org/10.1145/2872427.2883041}.

\bibitem[Tarjan(1975)]{Tarjan:1975:EGB:321879.321884}
R.~E. Tarjan.
\newblock {Efficiency of a Good But Not Linear Set Union Algorithm}.
\newblock \emph{Journal of the ACM (JACM)}, 22\penalty0 (2):\penalty0 215--225,
  4 1975.
\newblock ISSN 0004-5411.
\newblock \doi{10.1145/321879.321884}.
\newblock URL \url{http://doi.acm.org/10.1145/321879.321884}.

\bibitem[Tarjan and van Leeuwen(1984)]{Tarjan:1984:WAS:62.2160}
R.~E. Tarjan and J.~van Leeuwen.
\newblock {Worst-case Analysis of Set Union Algorithms}.
\newblock \emph{Journal of the ACM (JACM)}, 31\penalty0 (2):\penalty0 245--281,
  3 1984.
\newblock ISSN 0004-5411.
\newblock \doi{10.1145/62.2160}.
\newblock URL \url{http://doi.acm.org/10.1145/62.2160}.

\bibitem[Tarlow et~al.(2013)Tarlow, Swersky, Charlin, Sutskever, and
  Zemel]{pmlr-v28-tarlow13}
D.~Tarlow, K.~Swersky, L.~Charlin, I.~Sutskever, and R.~Zemel.
\newblock {Stochastic k-Neighborhood Selection for Supervised and Unsupervised
  Learning}.
\newblock In S.~Dasgupta and D.~McAllester, editors, \emph{Proceedings of the
  30th International Conference on Machine Learning}, volume~28 of
  \emph{Proceedings of Machine Learning Research}, pages 199--207, Atlanta,
  Georgia, USA, 2013. PMLR.
\newblock URL \url{http://proceedings.mlr.press/v28/tarlow13.html}.

\bibitem[Uhlmann(1991{\natexlab{a}})]{Uhlmann1991}
J.~K. Uhlmann.
\newblock {Satisfying general proximity / similarity queries with metric
  trees}.
\newblock \emph{Information Processing Letters}, 40\penalty0 (4):\penalty0
  175--179, 11 1991{\natexlab{a}}.
\newblock ISSN 00200190.
\newblock \doi{10.1016/0020-0190(91)90074-R}.
\newblock URL
  \url{http://linkinghub.elsevier.com/retrieve/pii/002001909190074R}.

\bibitem[Uhlmann(1991{\natexlab{b}})]{Uhlmann1991a}
J.~K. Uhlmann.
\newblock {Implementing Metric Trees to Satisfy General Proximity / Similarity
  Queries}.
\newblock Technical report, Naval Research Laboratory, Washington, D.C.,
  1991{\natexlab{b}}.

\bibitem[van~der Maaten(2014)]{VanderMaaten2014}
L.~van~der Maaten.
\newblock {Accelerating t-SNE using Tree-Based Algorithms}.
\newblock \emph{Journal of Machine Learning Research}, 15:\penalty0 3221--3245,
  2014.
\newblock URL \url{http://jmlr.org/papers/v15/vandermaaten14a.html}.

\bibitem[Walenstein et~al.(2007)Walenstein, Venable, Hayes, Thompson, and
  Lakhotia]{walenstein2007exploiting}
A.~Walenstein, M.~Venable, M.~Hayes, C.~Thompson, and A.~Lakhotia.
\newblock {Exploiting similarity between variants to defeat malware}.
\newblock In \emph{Proc. BlackHat DC Conf}, 2007.

\bibitem[Welford(1962)]{Welford1962a}
B.~P. Welford.
\newblock {Note on a Method for Calculating Corrected Sums of Squares and
  Products}.
\newblock \emph{Technometrics}, 4\penalty0 (3):\penalty0 419, 8 1962.
\newblock ISSN 00401706.
\newblock \doi{10.2307/1266577}.
\newblock URL \url{http://www.jstor.org/stable/1266577?origin=crossref}.

\bibitem[Yianilos(1993)]{Yianilos1993}
P.~Yianilos.
\newblock {Data structures and algorithms for nearest neighbor search in
  general metric spaces}.
\newblock In \emph{Proceedings of the fourth annual ACM-SIAM Symposium on
  Discrete algorithms}, page 311–321. Society for Industrial and Applied
  Mathematics, 1993.
\newblock URL \url{http://dl.acm.org/citation.cfm?id=313789}.

\end{thebibliography}

\appendix

\section{Corrections to Simplified Cover Tree} \label{sec:cover_correction}

We encountered two difficulties in replicating the simplified cover tree results of \textcite{Izbicki2015}. We detail these two issues and their remediations in this section for completeness and reproducibility. In the below algorithm descriptions we will use the same terminology and description as the algorithm's original paper, but note our changes in \textcolor{OliveGreen}{green}. 

We now review some of the properties needed to understand our corrections. The simplistic such property is that each node $p$ in the Cover tree has an associated level $l$, which we can obtain as $l = \text{level}(p)$. Each child $c_p$ of $p$ must also satisfy the property that $\text{level}(p) = \text{level}(c_p)+1$. 

Using a node's level, we can define its \textit{coverdist} as $\text{coverdist(p)} = 2^{\text{level}(p)}$. Each child $c_p$ of $p$ will satisfy the covering invariant property, $d(c_p, p) \leq \text{coverdist}(p), \forall c_p \in \text{children}(p)$.

We also must make use of the \textit{maxdist} bound discussed in \autoref{sec:cover_tree}, which we make more explicit as: $\text{maxdist}(p) = \argmax_{d_p \in \text{descendants}(p)} d(d_p, p)$. This is the maximum distance from one node $p$ to \textit{any} descendant note of $p$. If $p$ is a leaf node, meaning it has no children, then $\text{maxdist}(p) = 0$. 

\subsection{Nearest Neighbor Correction}

We present the revised nearest neighbor search procedure for the simplified Cover-tree in \autoref{algo:cover_search}. The green $d(x, p)$ term was originally presented to be $d(y, q)$. We show that this is not correct using a simple counter example using scalar node values and the euclidean distance. 

\begin{algorithm}[!htb]
\caption{Cover Tree Find Nearest Neighbor}
\label{algo:cover_search}
\begin{algorithmic}[1]
\Require cover tree $p$, query point $x$, nearest neighbor so far $y$
\If { $d(p,x)<d(y,x)$}
	\State $y \gets p$
\EndIf
\For{each child $q$ of $p$ sorted by distance to $x$}
	\If {$d(y,x)>$ \textcolor{OliveGreen}{$d(x,q)$} $-\text{maxdist}(q)$} \Comment{Original paper used $d(y,q)$}
    	\State $y \gets \text{findNearestNeighbor}(q,x,y)$
    \EndIf
\EndFor
\State \Return $y$
\end{algorithmic}
\end{algorithm}  

Consider the Cover-tree  with root $\alpha$, that stores value 5. $\alpha$ has one child, $\beta$, which has the value $-2$. This is the whole tree. 

We would begin on line one of the algorithm, with $p \gets \alpha$ and we will use our query point $x$ to have a value of $0$. $d(p, x)$ is 5, and we have no nearest neighbor so far, so $y \gets p$ (which is $\alpha$) becomes the nearest neighbor so far. 

We will obtain $q \gets \beta$ as it is the only child of $\alpha$, which leads us to evaluate the original expression 

$$\underbrace{d(y, x)}_{=5-0=5} > \underbrace{d(y, q)}_{=5-(-2)=7} - \underbrace{\text{maxdist}(q)}_{=0}$$

Because $5 > 7$ is false, the if statement fails, and we then break from the loop, returning $y$ as the nearest neighbor to $x$ with a distance of 5. But $x$'s value is 0, and $\beta$'s is $-2$, which is a distance of only two away.

\subsection{Insertion Correction}

We also provide a correction to the insertion procedure of the simplified Cover-tree. Our fixed version is presented in \autoref{algo:cover_build}, with the green text indicating only added statements to the algorithm. 

The issue with the original procedure occurs when an outlier $x$ is inserted into the index, the distance from which to any point in the dataset is larger than the largest pairwise distance of any two points in the existing Cover-tree. This is because the $2 \text{coverdist}(p) \geq maxdist(p)$ in all cases. If $x$ is farther than the maximum pairwise distance, then the simple bound on line four may be true for a all points in a valid cover tree. This means the loop will never exit, and will simply continue re-structuring the tree in search of a non-existing node that can satisfy the loop condition. 

We fix this by keeping track of the points visited in the tree, and only loop while there is a potential candidate remaining. If no such candidate occurs because we have visited all possible leaf nodes, the loop must exit so that the outlier may be inserted as the new root of the tree. 

\begin{algorithm}[!htb]
\caption{Simplified Cover Tree Insertion}
\label{algo:cover_build}
\begin{algorithmic}[1]
\Require Query $q$, desired number of neighbors $k$
\Procedure{insert}{cover tree $p$, data point $x$}
\If {$d(p,x) > \text{covdist}(p)$}
	\State \textcolor{OliveGreen}{$z \gets \emptyset$}
	\While {$d(p,x)>2 \text{covdist}(p)$ \textcolor{OliveGreen}{\textbf{and} $|\text{descendants}(p)| > |z|$} }
    	\State Remove any leaf $q$ from $p$\textcolor{OliveGreen}{$\setminus z$}
        \State $p' \gets $ tree with root $q$ and $p$ as only child
        \State $p \gets p'$
    \EndWhile
    \State \Return tree with $x$ as root and $p$ as only child
\EndIf
\State \Return $\text{INSERT\_}(p,x)$

\EndProcedure
\Procedure{insert\_}{cover tree $p$, data point $x$}
\ForAll{$q \in \text{children}(p)$}
	\If {$d(q,x)\leq \text{covdist}(q)$}
    	\State $q' \gets \text{INSERT\_}(q,x)$
        \State $p' \gets p \text{ with child } q \text{ replaced with } q'$
        \State \Return $p'$
    \EndIf
\EndFor
\State \Return $p$ with $x$ added as a child
\EndProcedure
\end{algorithmic}
\end{algorithm}  

From a practical implementation perspective, we note two additional choices. First, rather than attempt to remove leaf nodes in the specified form above, it is easier to define a specific leaf removal order and leaf insertion order. For example, if one always removes the least recently added leaf node, we will obtain a consistent ordering of the leaf nodes as we iterate line four of the algorithm. This makes it easy to use simple cycle detection to determine that the all possible children have been visited, and then escape the loop when this occurs. 

To speed up insertion of outlier points, we also note that the covering invariant can be used to catch extreme outliers. If $d(p, x) > 4 \text{covdist}(p)$, then we can skip the loop entirely and proceed directly to line eight of the algorithm. This bound is easy to see, as $ 2 \text{covdist}(p) \geq \text{maxdist}(p)$. Assuming that there exists a descendant point $\gamma$ that is maximally far from $p$. Let $\zeta$ be the point maximally far from $\gamma$, and let $d(\gamma, \zeta)$ be the maximal pairwise distance for all points in the Cover-tree. Direct application of the triangle inequality gives us 

$$d(\gamma, \zeta) \leq d(\gamma, p) + d(p, \zeta)$$

This bounds the distance between these points by their distance to the root. The covering invariant tells us that $2 \text{coverdist}(p) \geq maxdist(p)$. Therefore it must be the case that

$$d(\gamma, \zeta) \leq 2 \text{coverdist}(p) + 2 \text{coverdist}(p)$$

Which reduces to the bound $d(\gamma, \zeta) \leq 4 \text{coverdist}(p)$. Thus if a new query violates this bound, we know that no point in the whole tree can satisfy the loop on line 4.

\end{document}